\documentclass[11pt, a4paper]{article}
\usepackage[utf8]{inputenc}

\usepackage[style=nature, maxbibnames=99, sorting=none]{biblatex}
\usepackage[margin=1in]{geometry}
\usepackage{bm}
\usepackage{setspace}
\usepackage{amsmath}
\usepackage{graphicx}
\usepackage{arydshln}
\usepackage{rotating}
\usepackage{hyperref}
\usepackage{caption}
\usepackage{authblk}
\usepackage{adjustbox}
\graphicspath{ {./plots/} }

\newcommand\Nbound{N_{\text{bound}}}

\title{\textbf{Online error control for platform trials}}

\author[a]{David S.\ Robertson\footnote{david.robertson@mrc-bsu.cam.ac.uk}}
\author[b]{James M.\ S.\ Wason}
\author[c]{Franz K\"{o}nig}
\author[c]{Martin Posch}
\author[a]{Thomas Jaki}

\affil[a]{MRC Biostatistics Unit, School of Clinical Medicine, University of Cambridge, UK \vspace{6pt}}

\affil[b]{Population Health Sciences Institute, 
Faculty of Medical Sciences, Newcastle University, UK \vspace{6pt}}

\affil[c]{Section of Medical Statistics, Medical University of Vienna, Austria}

\date{\vspace{-18pt}}

\begin{document}

\maketitle

\onehalfspace

\begin{abstract}
Platform trials evaluate multiple experimental treatments under a single master protocol, where new treatment arms are added to the trial over time. Given the multiple treatment comparisons, there is the potential for inflation of the overall type~I error rate, which is complicated by the fact that the hypotheses are tested at different times and are not all necessarily pre-specified. Online error control methodology provides a possible solution to the problem of multiplicity for platform trials where a relatively large number of hypotheses are expected to be tested over time. In the online testing framework, hypotheses are tested in a sequential manner, where at each time-step an analyst decides whether to reject the current null hypothesis without knowledge of future tests but based solely on past decisions. Methodology has recently been developed for online control of the false discovery rate as well as the familywise error rate (FWER). In this paper, we describe how to apply online error control to the platform trial setting, present extensive simulation results, and give some recommendations for the use of this new methodology in practice. We show that the algorithms for online error rate control can have a substantially lower FWER than uncorrected testing, while still achieving noticeable gains in power when compared with the use of a Bonferroni procedure. We also illustrate how online error control would have impacted a currently ongoing platform trial. \\[12pt]
\end{abstract}


\section{Introduction}


There is a strong need to conduct high-quality evaluations of new interventions aimed at improving health of patients. The highest quality of evidence comes from Randomised Controlled Trials (RCTs). However, the cost of RCTs is high and increasing, leading to the very high costs of bringing new drugs to market~\cite{dimasi2016innovation} and evaluation of other types of intervention~\cite{bentley2019conducting}. This has led to focus on methods that can improve the operational and statistical efficiency of conducting clinical trials.

One important class of efficient approaches are platform trials~\cite{saville2016efficiencies,park2020overview, meyer2020evolution}. Platform trials set up an infrastructure that allows evaluation of multiple intervention arms under a single master protocol. Platform trials may use an adaptive design to drop non-promising intervention arms early and can add new interventions as they become available for evaluation. Some interventions can also be evaluated only within particular patient subgroups if warranted. These properties all can lead to increased efficiency~\cite{saville2016efficiencies, parmar2014more, pallmann2018adaptive}. However, they also introduce challenges, both operational~\cite{hague2019changing} and statistical~\cite{lee2021statistical}.

One statistical challenge, given the multiple treatment comparisons, is control of type~I error rates. Although the need for formal control of type~I error rates in trials of distinct treatments is controversial~\cite{wason2014correcting, howard2018recommendations,collignon2020current,collignon2021collaborative,wason2021controlling,parker2020non, bretz2020commentary}, it may be required in some regulatory settings and desirable in other situations where the treatments are related in some way. In a platform trial, controlling type~I error rates is complicated by the fact that the hypotheses are tested at different times and are not all pre-specified.

Online error control methodology provides a possible solution to the problem of multiplicity for platform trials where multiple treatments are tested over time~\cite{Robertson2018}. In the online testing framework, hypotheses are tested in a sequential manner, where at each time-step a decision is made whether to reject the current null hypothesis without knowledge of future tests but based solely on past decisions. Methodology has recently been developed for online control of the false discovery rate (FDR) \cite{Javanmard2018, Ramdas2017, Ramdas2018, Tian2019a, Zrnic2018, Zrnic2019} as well as the familywise error rate (FWER)~\cite{Tian2019}, so that the relevant error rate is controlled at all times throughout the trial. 

In order to apply online error control methodology to platform trials and evaluate its performance, there are some general features specific to this setting that need to be explored. Firstly, one key consideration is the total number of hypothesis tests that is reasonable to envisage for a platform trial. Thus far, most of the literature on online testing has tended to focus on applications with a very large ($\geq 1000$) number of hypotheses. Secondly, many existing algorithms~\cite{Javanmard2018, Ramdas2017, Ramdas2018, Tian2019, Tian2019a} for online FDR or FWER control assume independence between the $p$-values (or equivalently, the test statistics). However, in the platform trial setting there is a positive dependence induced by shared control arm information, and so potential inflation of the relevant type~I error rate needs to be considered. Lastly, in the clinical trial setting the usual error rate considered (at least for confirmatory trials) is the FWER~\cite{EMAmultiplicity, FDAmultiplicity}. Hence, it would be useful to see how the FWER is impacted when using methods that control the FDR.

Our aim in this paper is to demonstrate the advantages and disadvantages of using online error control methodology in the platform trial setting through an extensive simulation study looking at varying numbers of treatment arms, patterns of expected treatment responses and arm entry times, as well as assumptions about the upper bound on the total number of experimental treatments to be tested.

In Section~\ref{sec:alg_online}, we introduce the different algorithms for online error control that we will explore, and then in Section~\ref{sec:simul_study} we describe the idealised platform trial set-up that we use for the simulations and the different simulation scenarios. Section~\ref{sec:simul_results} provides the simulation results, focusing on fully sequential testing (Section~\ref{subsec:fully_seq}) as well as testing treatment arms in batches (Section~\ref{subsec:batched}). In Section~\ref{sec:case_study} we present a case study based on the STAMPEDE platform trial~\cite{james2008stampede} of therapies for prostate cancer. We conclude with recommendations and discussion in Section~\ref{sec:discuss}.

\section{Algorithms for online error control}
\label{sec:alg_online}

\begin{figure}[ht!]
\centering
\includegraphics[width=0.7\textwidth]{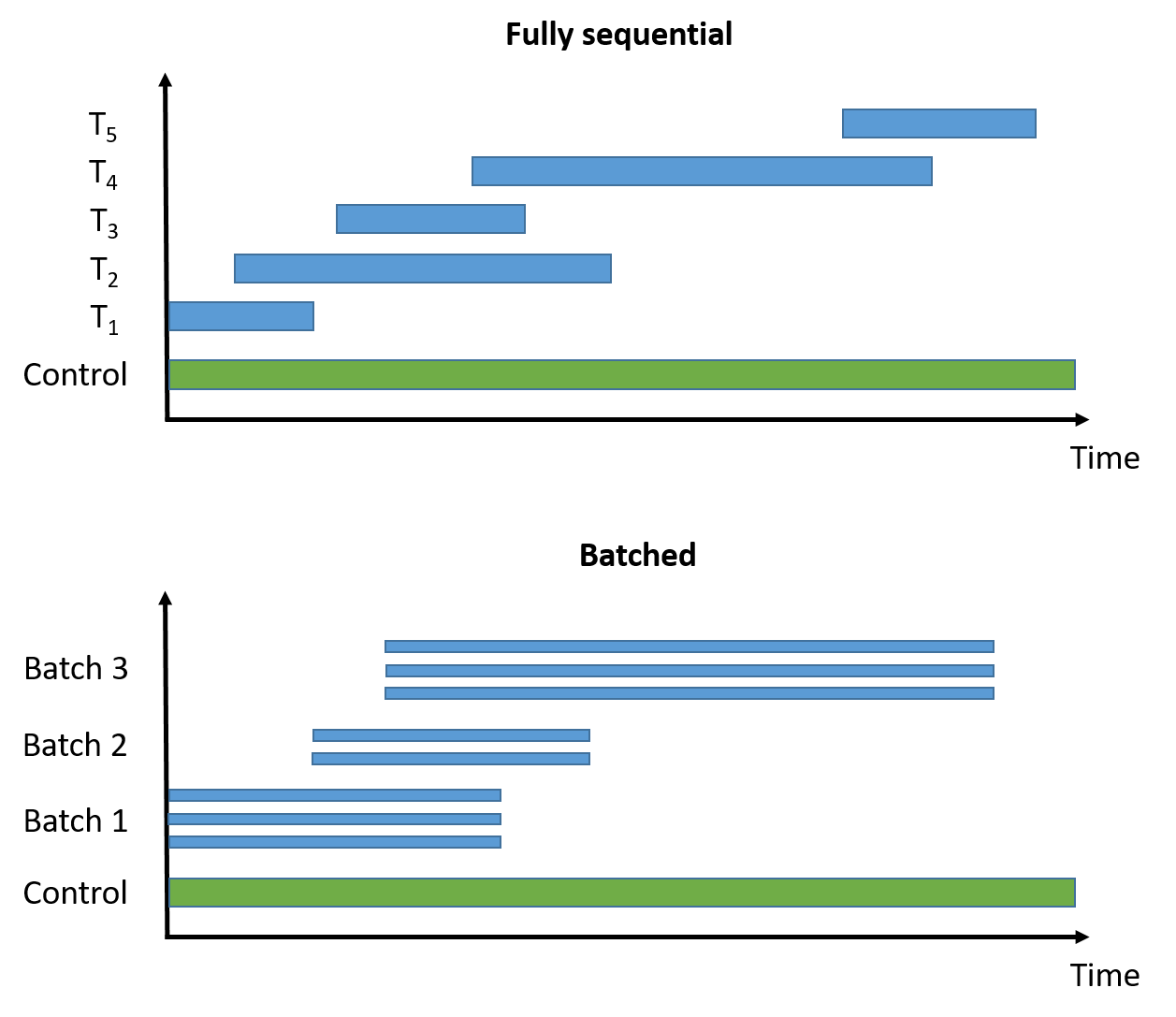}
\caption{Illustration of fully sequential and batched testing for a platform trial testing experimental treatments $T_i$ against a common control. For the online testing procedures, a hypothesis test can be conducted whenever new results become available (i.e.\
whichever treatment arm is the $i$-th one to be analysed is tested at level $\alpha_i$). For offline methods (e.g.\ the Benjamini-Hochberg procedure), testing can only occur when all the trial data are available. }
\label{fig:seq_batch}
\end{figure}

We now briefly introduce the online testing algorithms that we consider, classified into those for fully sequential testing (i.e.\ where each hypothesis is tested one after another) and those for batched testing (i.e.\ where groups of hypotheses are available to be tested at the same time). An illustration of the difference is given in Figure~\ref{fig:seq_batch}. All of the algorithms require a pre-specified significance level $\alpha$, as well as a sequence of non-negative numbers $\gamma_i$ that sum to 1. We provide references giving further details of the algorithms for the interested reader. The algorithms described below are also available to use via the \texttt{onlineFDR} R package~\cite{robertson2019onlineFDR}.

\subsection{Fully sequential online testing}

The algorithms for fully sequential online testing are used to calculated the adjusted testing levels $\alpha_i$ for testing the null hypothesis $H_i$, i.e.\ $H_i$ is rejected if the associated $p$-value $p_i$ satisfies $p_i \leq \alpha_i$. 

\begin{itemize}
    \item \textit{LOND}: One of the first online testing algorithms proposed, LOND (Levels based on Number of Discoveries) provably controls the FDR for independent~\cite{Javanmard2018} and positively dependent $p$-values~\cite{Zrnic2018}. In LOND, the $\alpha_i$ are calculated by simply multiplying the sequence~$\alpha \gamma_i$ by the number of discoveries (i.e.\ the number of rejected hypotheses) that have been made thus far.
    
    \item \textit{LORD}: Another early online testing algorithm, LORD (Levels based on Recent Discovery) provably controls the FDR for independent $p$-values~\cite{Javanmard2018, Ramdas2017}. LORD takes advantage of so-called ``alpha-investing'', in which hypothesis tests costs some amount from the error budget (or ``alpha-wealth''), but a discovery earns some of the error budget back. The $\alpha_i$ in LORD depend not only on how many discoveries have been made but also on the timing of these discoveries. The more recent discoveries there are, the higher the alpha-wealth will be.
    
    \item \textit{SAFFRON}: Proposed as an improvement to LORD, the SAFFRON (Serial estimate of the Alpha Fraction that is Futilely Rationed On true Null hypotheses) algorithm provably controls the FDR for independent $p$-values~\cite{Ramdas2018}. Intuitively, SAFFRON focuses on the stronger signals in an experiment (i.e.\ the smaller $p$-values). By never rejecting weaker signals (i.e.\ larger $p$-values), SAFFRON preserves alpha-wealth. When a substantial fraction of null-hypotheses are false, SAFFRON will often be more powerful than LORD. 
    
    \item \textit{ADDIS}: Proposed as an additional improvement to both LORD and SAFFRON, the ADDIS (ADaptive algorithm that DIScards conservative nulls) algorithm provably controls the FDR for independent $p$-values~\cite{Tian2019}. Intuitively, ADDIS builds on the SAFFRON algorithm by potentially investing alpha-wealth more effectively through the discarding of the weakest signals (i.e.\ the largest $p$-values) in a principled way. This procedure can gain appreciable power if the null $p$-values are conservative, i.e.\ stochastically larger than the uniform distribution.
    
    \item \textit{ADDIS-spending}: Unlike all of the algorithms mentioned above, ADDIS-spending provably controls the FWER (in the strong sense) for independent $p$-values~\cite{Tian2019a}. ADDIS-spending shares similarities with ADDIS in setting thresholds based on the size of the $p$-values for hypotheses to be selected for testing.
    
\end{itemize}

\subsection{Batched online testing}

The algorithms for batched testing use well-known offline procedures for FDR control (i.e.\ procedures that require all of the $p$-values to be known before testing) for each batch, in such a way that the FDR is controlled across all of the batches as well. Three related algorithms were proposed by Zrnic et al.~\cite{Zrnic2019}:

\begin{itemize}
    \item \textit{BatchBH}: this algorithm provably controls the FDR when the $p$-values are independent within and across batches. BatchBH runs the Benjamini-Hochberg (BH) procedure on each batch, where the values of $\alpha_i$ depend on the number of previous discoveries.
    
    \item \textit{BatchPRDS}: this algorithm is a modification of BatchBH, which provably controls the FDR when the p-values in one batch are positively dependent, and independent across batches.
    
    \item \textit{BatchStBH}: provably controls the FDR when the $p$-values are independent within and across batches. BatchStBH runs the Storey Benjamini-Hochberg (StBH) procedure on each batch, where the values of $\alpha_i$ depend on the number of previous discoveries. The StBH procedure can potentially make more rejections than BH by adapting to the (estimated) number of non-nulls.
    
\end{itemize}


\section{Simulation study}
\label{sec:simul_study}

We now describe the platform trial set-up (Section~\ref{subsec:platform_setup}) used in our simulation study, as well as the simulation scenarios (Section~\ref{subsec:sim_scenarios}) that we explore.

\subsection{Platform trial set-up}
\label{subsec:platform_setup}

Consider an idealised platform trial that eventually tests a total of~$K$ experimental treatments $T_1, \ldots, T_K$ against a common control~$T_0$ (see Figure~\ref{fig:STAMPEDE} for a graphical example). Let $\mu_i$ denote the expected response for treatment~$T_i$, and define the treatment effect $\theta_i = \mu_i - \mu_0$ for $i \in \{1, \ldots, K\}$. The null hypotheses of interest are given by $H_{0i}: \theta_i \leq 0$ with corresponding alternatives $H_{1i} : \theta_i > 0$, which we test at error level~$\alpha$ (for either FWER or FDR). Let $n_i$ denote the pre-specified sample size for each experimental treatment, and for simplicity assume that an equal number of patients are eventually allocated to each experimental arm (i.e.\ $n_1 = \cdots = n_K = n$). 
We use the simplifying assumption that an experimental treatment continues in the trial until~$n$ patients have been allocated to that arm and their outcomes observed. The experimental treatment is then formally tested for effectiveness by comparing with \textit{concurrent} controls (i.e.\ only the outcomes from patients allocated to the control group during the time that the experimental treatment was active in the trial). We return to the issue of conducting interim analyses in Section~\ref{sec:discuss}. 

We assume that the observation $X_{ij}$ from patient~$j$ on treatment~$T_i$ is distributed (at least asymptotically) as $X_{ij} \sim N(\mu_i, \sigma^2)$, where $\sigma$ is known. The mean response for patients on experimental treatment~$T_i$ ($i > 0$) is then given by $\bar{X}_i = \frac{1}{n} \sum_{j=1}^n X_{ij}$, with corresponding concurrent control mean $\bar{X}_{0}^{(i)} = \frac{1}{N_0(i)} \sum_{j \in \mathcal{T}_i} X_{0j}$, where $N_0(i)$ and $\mathcal{T}_i$ denote the number and index set, respectively, of the control observations that are concurrent with~$T_i$.
The comparisons between the experimental treatments and the control are based on normally-distributed test-statistics \[
Z_i = \frac{\bar{X}_i - \bar{X}_0^{(i)}}{\sigma \sqrt{\frac{1}{N_0(i)} + \frac{1}{n}}} \sim N\!\left(\frac{\theta_i}{\sigma \sqrt{\frac{1}{N_0(i)} + \frac{1}{n}}}, 1\right)
\] where $\bar{X}_i - \bar{X}_0$ is the observed difference in means. The corresponding one-sided $p$-value for treatment~$T_i$ is given by $p_i = 1 - \Phi(Z)$, where $\Phi$ is the standard normal cdf.

We assume for simplicity that experimental treatment~$T_i$ enters the trial at time~$t_i$, and remains in the trial until time $(t_i + r)$. The platform trial starts at time~0 and finishes once all~$K$ experimental treatments have been tested. Hence the temporal structure of the platform trial can be described by the value of~$r$ and the vector $\bm{t} = (t_1, \ldots, t_K)$. Without loss of generality, we impose the restriction that $t_1 \leq \cdots \leq t_K$.
We assume that $n/r$ patients are allocated to each arm (including the control) per unit time, where~$r$ is chosen so that $n/r$ is an integer.

\subsection{Simulation scenarios}
\label{subsec:sim_scenarios}

In our simulation study, we consider $K \in \{5, 10, 15, 20\}$, and set $\alpha = 0.025$, $n = 50$, $\mu_0 = 0$, $\sigma = 1$ and $r = 10$. Table~\ref{tab:sim_overview} gives an overview of the scenarios, algorithms and performance metrics used in the simulation study, with further details given below.

\begin{table}[ht!]
    \centering
    \renewcommand{\arraystretch}{1.5}
    \begin{adjustbox}{max width=\textwidth}
    \begin{tabular}{ l l l l }
    \textbf{Treatment means} & \textbf{Arm entry times} & \textbf{Algorithms} & \textbf{Performance metrics} \\ \hline
    Global null & All arms at once & Bonferroni, ADDIS-spending & FWER and FDR\\
    Fixed means & Batches of size 5 &  ADDIS, LORD, SAFFRON, BH  & Disjunctive power \\
    Staircase scenario & Staggered starts & BatchBH, BatchPRDS, BatchStBH & Sensitivity \\
    & Fully sequential &  \\ \hline
    \end{tabular}
    \end{adjustbox}
    \caption{Overview of simulation scenarios and performance metrics.}
    \label{tab:sim_overview}
\end{table}

\noindent We consider the following scenarios for the treatment means~$\mu_i$, $i \in \{1, \ldots, K\}$:
\begin{enumerate}
    \item Global null: $\mu_1 = \cdots = \mu_K = 0$

    \item Fixed means: $m$ effective treatments which are either all tested at the start or the end of the platform trial, or are tested in a random order. We set $\mu_i = 0.5$ for $i \in I$ and $\mu_i = 0$ otherwise, where \begin{itemize}
        \item[] $I = \{i : 1 \leq i \leq m\}$ (Early)
        \item[] $I = \{i : K - m + 1 \leq i \leq K\}$ (Late)
        \item[] $I$ is a set of $m$ values drawn randomly from $\{1, \ldots, K\}$ without replacement (Random)
    \end{itemize} for $m \in \{ 1, K/5+1, 2K/5+1 \}$.
    
    \item Staircase scenario:
        \begin{itemize}
            \item[] $\mu_i = (i-\lceil{K/2}\rceil)/K$ (Decreasing)
            \item[] $\mu_i = (\lceil{K/2}\rceil-i+1)/K$ (Increasing)
            \item[] $\mu_i$ is drawn randomly from the set $\{(i-\lceil{K/2}\rceil)/K : 1 \leq i \leq K\}$ without replacement (Random)
        \end{itemize}
    
    
\end{enumerate}

\noindent We also consider the following patterns of arm entry times~$\bm{t}$:
\begin{enumerate}
    \item All arms at once (offline testing): $t_i = 0$
    
    \item Batches of size $b$: $t_i = r(j-1)$ for $i \in \{b(j-1)+1, \ldots, bj \}$ and $j \in \{ 1, \ldots, K/b\}$, where $b = 5$
    
    \item Staggered starts: $t_i = r(i-1)/s$ for $s \in \{2, 5\}$
    
    \item Fully sequential: $t_i = r(i-1)$ \\[-6pt]
    
\end{enumerate}

\noindent Finally, we explore different assumptions about the upper bound, $\Nbound$, on the total number of experimental treatments to be tested. We consider $\Nbound \in \{K, 2K, 5K\}$, i.e.\ considering the impact of more conservative choices of $\Nbound$. Note that $\Nbound$ should be chosen before the start of the platform trial. We return to the issue of what happens if $\Nbound < K$ in Section~\ref{sec:discuss}.

For each of the simulation scenarios described above, we compare the following algorithms for FWER and FDR control:
\begin{itemize}
    \item[] \textbf{FWER}: Bonferroni and ADDIS-spending.
    
    \item[] \textbf{FDR}: ADDIS, SAFFRON and LOND. For batched entry times, we consider the BatchBH, BatchPRDS and BatchStBH procedures. We also use the standard Benjamini-Hochberg (BH) procedure as a comparator. \\[-6pt]
    
\end{itemize}

We give more details about the exact implementation of these algorithms in Section~A of the Supporting web materials. Note that we are assuming that Bonferroni tests each hypothesis at level~$\alpha/\Nbound$ (as opposed to $\alpha/K$), and that the BH procedure is an offline procedure that requires all hypotheses to be tested at once.

For each simulation scenario, we report the following performance measures:
\begin{itemize}
    \item[] \textbf{Type I error rates}: FWER and FDR
    
    \item[] \textbf{Power metrics}: Disjunctive power (probability of rejecting at least one non-null hypothesis) and Sensitivity (proportion of non-null hypotheses that are rejected) \\[-6pt]
    
\end{itemize}



\section{Simulation results}
\label{sec:simul_results}

To start with, we consider the type~I error rate of the different algorithms, in particular checking whether the FDR is controlled. This is of interest because apart from LOND, all of the online testing algorithms only provably control the FDR under independence of the $p$-values.  Figure~B1 in Section~B.1 of the Supporting web materials shows the FDR under the global null (and hence the FWER = FDR), for varying patterns of arm entry times and $\Nbound = K$. Note that the global null maximises the FWER for multi-arm multi-stage (MAMS) designs~\cite{magirr2012generalized}. Using uncorrected testing leads to a highly inflated FDR/FWER above the nominal 2.5\%, which can be as high as 40\% when $K = 20$. In contrast, the online testing algorithms have a FDR/FWER which is equal to or below the nominal level. The one exception is SAFFRON, which has a slight inflation of the FDR/FWER to around 3\%. 

\subsection{Fully sequential setting}
\label{subsec:fully_seq}

Continuing with the FDR and now focusing on the fully sequential setting, Figure~B2 in Section~B.2 of the Supporting web materials shows the worst and best case FDR (corresponding to all the effective treatments appearing Early and Late, respectively) for fixed means and different numbers of effective treatments, with $\Nbound = K$. Again we see how uncorrected testing can lead to a highly inflated FDR, although this inflation reduces as the number of effective treatments increases. This time, all the online testing procedures control the FDR at or below the nominal 2.5\%, but they can be noticeably conservative. This is the case for LORD, as well as ADDIS-spending for $(K/5+1)$ and $(2K/5+1)$ effective treatments. Note though that the Bonferroni procedure is also highly conservative in these settings.

\begin{figure}[ht]
    \centering
    \includegraphics[width = \textwidth]{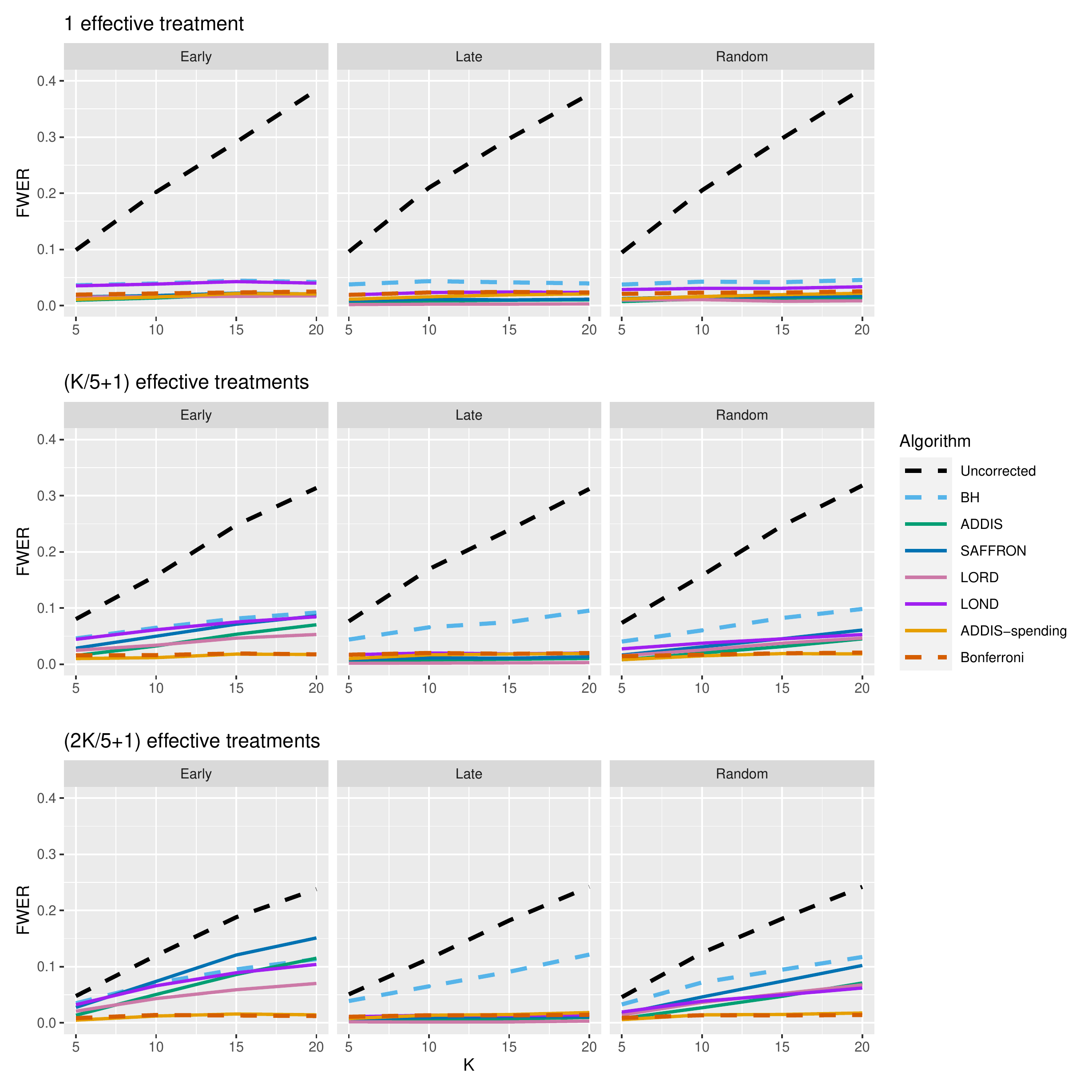}
    \caption{FWER for fixed means and different numbers of effective treatments, with $\Nbound = K$.}
    \label{fig:fully_seq_FWER}
\end{figure}

We now move on to examining the FWER for the different procedures. Of course, the FDR-controlling algorithms would not be expected to also control the FWER in general, but it is of interest to see how large any inflation in the FWER can be. In general, the magnitude of the inflation depends critically on the effect sizes chosen for the alternative hypotheses. Figure~\ref{fig:fully_seq_FWER} shows the FWER under the early and late scenarios, as well as the FWER under the random ordering of the effective treatments, all with $\Nbound = K$. When there is only 1 effective treatment the FWER is controlled by all the online testing algorithms, but (apart from LOND) they are even more conservative than Bonferroni. When there are $(K/5+1)$ effective treatments, the FWER is inflated above 2.5\% for all the online testing procedures (apart from ADDIS-spending) when $K > 5$ in the Early scenario and $K > 15$ in the Random scenario. The LOND algorithm has the largest inflation in the Early scenario, which is similar to the inflation seen using the BH procedure. Finally, when there are $(2K/5+1)$ effective treatments, there is FWER inflation for all the online algorithms (except for ADDIS-spending) when $K > 5$ in the Early scenario and $K > 10$ in the Random scenario. This time, SAFFRON has the highest inflation, which can be higher than BH in the Early scenario, although still noticeably lower than uncorrected testing. Note that in the random setting, for both $(K/5+1)$ and $(2K/5+1)$ effective treatments, when $K = 15$ the FWER of the online testing procedures (apart from SAFFRON) is controlled at~5\%.

\begin{figure}[ht!]
    \centering
    \includegraphics[width = \textwidth]{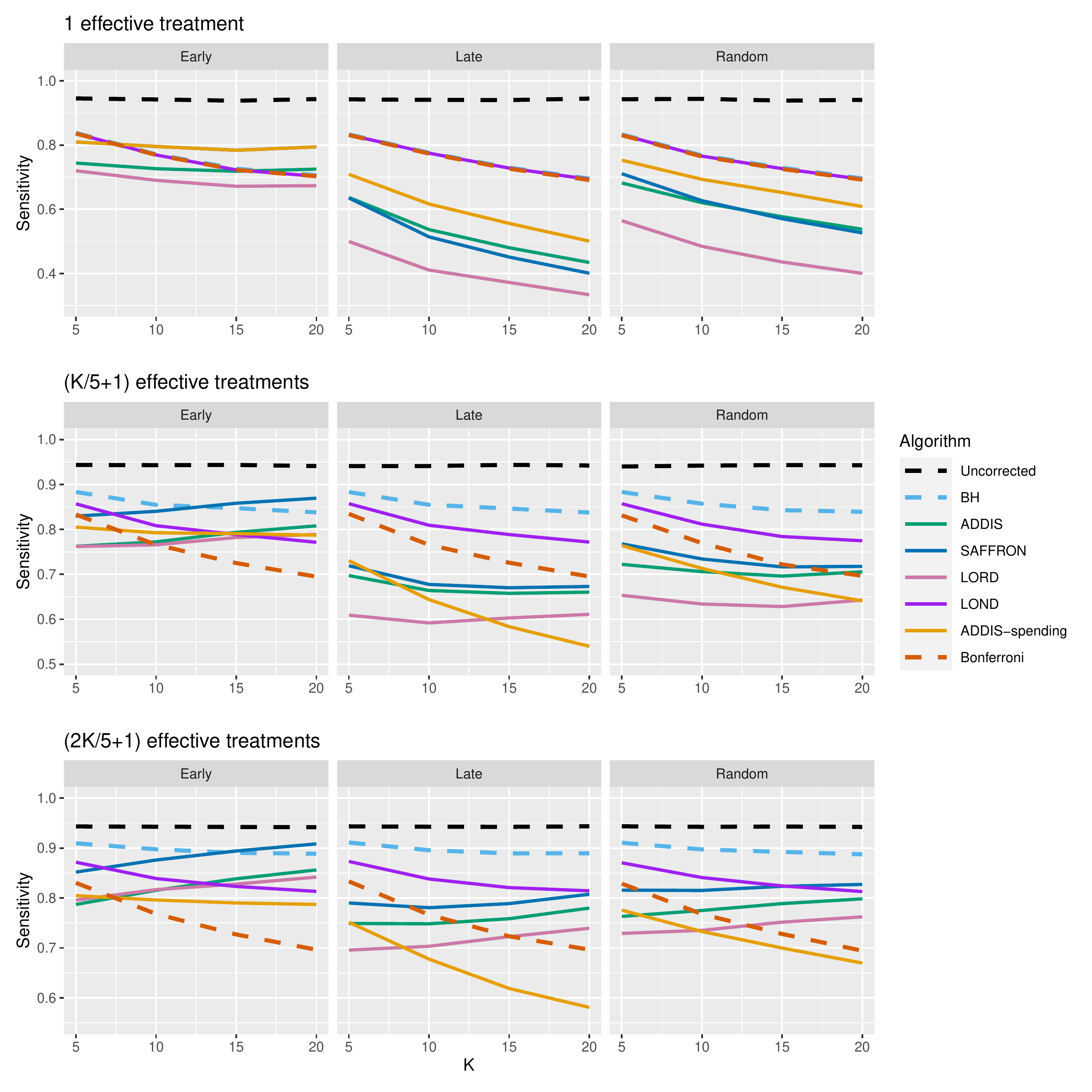}
    \caption{Sensitivity for fixed means and different numbers of effective treatments, with $\Nbound = K$. Note that the $y$-axis scales differ as the number of effective treatments vary for readability.}
    \label{fig:fully_seq_sens}
\end{figure}

The other side of the story is the power of the trial, and to start with we focus on sensitivity as a measure of the proportion of the truly effective treatments that are declared efficacious. Figure~\ref{fig:fully_seq_sens} shows the sensitivities for the Early, Late and Random scenarios, again with $\Nbound = K$. For ease of reference, we also include plots showing the Sensitivity and FWER together in Figures~B3--B5 in Section~B.2 of the supporting web materials.
With only~1 effective treatment, the sensitivity of all of the online algorithms apart from LOND is substantially lower than using Bonferroni. Even in the best case (Early scenario), only SAFFRON and ADDIS-spending have a higher sensitivity than Bonferroni when $K \geq 10$.
When there are $(K/5+1)$ effective treatments, again the sensitivities of all of the online algorithms apart from LOND are lower than Bonferroni in the Late and Random scenarios. LOND maintains a sensitivity roughly halfway that between the Bonferroni and BH procedures. In the best case (Early scenario), SAFFRON has a higher sensitivity than Bonferroni for $K > 5$ (and even BH for $K \geq 15$), while the other online algorithms also achieve a higher sensitivity for $K > 10$. 
Finally, with $(2K/5+1)$ effective treatments, we again see that LOND maintains a sensitivity roughly halfway that between the Bonferroni and BH procedures. Under the Random scenario, SAFFRON, ADDIS and LORD have a higher sensitivity than Bonferroni for $K>5$, $K>10$ and $K \geq 15$, respectively. In the worst case (Late scenario), the equivalent values of~$K$ are $K\geq 10$, $K>10$ and $K > 15$, respectively. Finally, in the best case (Early scenario), all the online testing algorithms outperform Bonferroni for $K \geq 10$.

\begin{figure}[ht!]
    \centering
    \includegraphics[width = \textwidth]{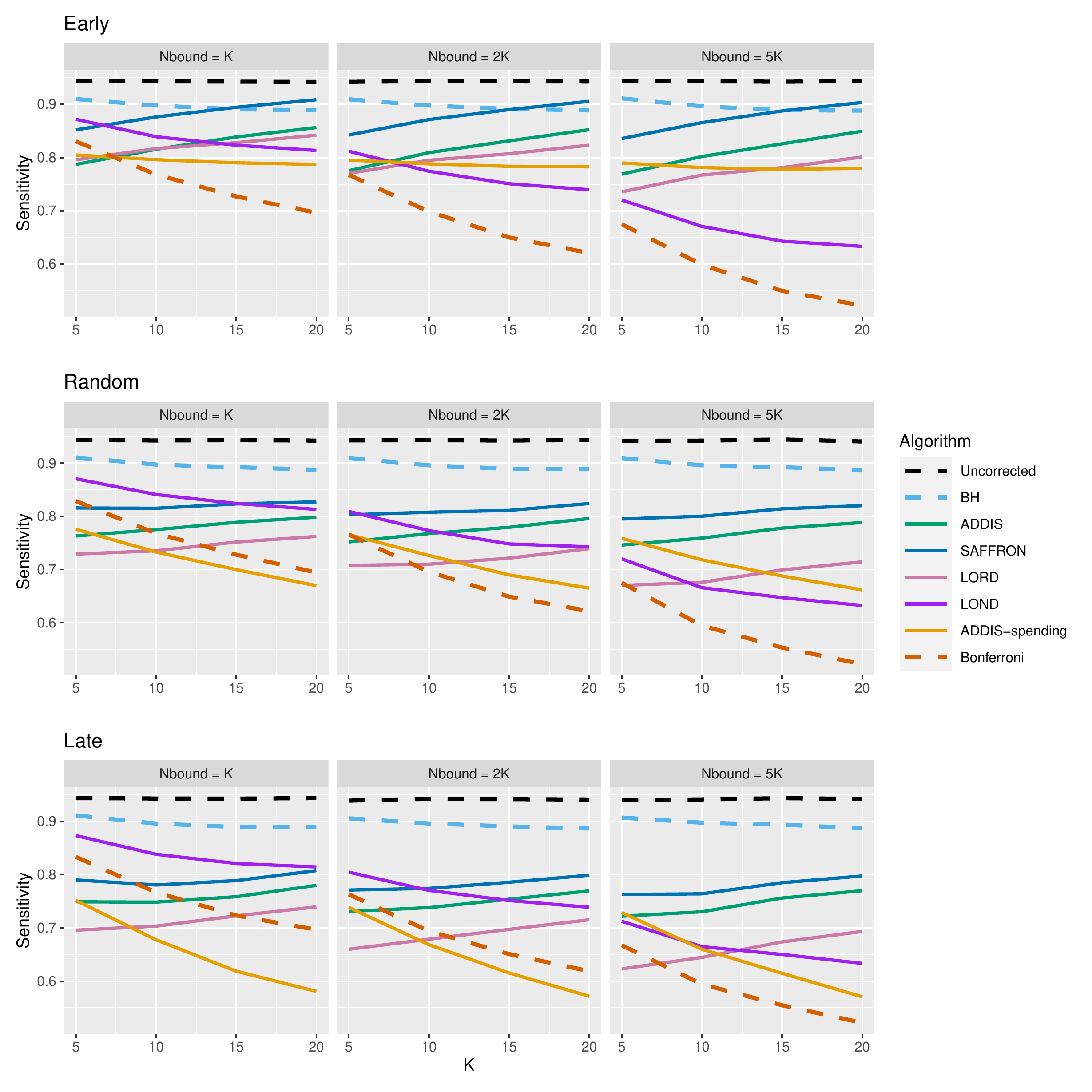}
    \caption{Sensitivity for fixed means and ($2K/5+1$) effective treatments, with varying $\Nbound$.}
    \label{fig:fully_seq_sens_Nbound}
\end{figure}

Thus far we have used an upper bound $\Nbound = K$, i.e.\ exactly equal to the number of treatments that are tested. In practice, this is unlikely to be known prior to the platform trial starting, and a more conservative (i.e.\ higher) value of $\Nbound$ would be used. Figure~\ref{fig:fully_seq_sens_Nbound} shows the effect of increasing the value of $\Nbound$ on the sensitivity of the algorithms, where there are $(2K/5+1)$ effective treatments. The key takeaway here is that as $\Nbound$ increases, the sensitivity of Bonferroni, LOND and LORD noticeably decreases, with Bonferroni having the largest relative decrease, whereas the sensitivity of all the other algorithms remains virtually unchanged. This means that there is a greater advantage in using online algorithms (in terms of sensitivity) compared with Bonferroni as $\Nbound$ increases. For example, in the extreme case where $\Nbound = 5K$, then under a random ordering of treatment means, all of the online algorithms outperform Bonferroni, with the relative advantage increasing with~$K$. The same is seen for the more realistic scenario of $\Nbound = 2K$, apart from LORD for $K < 10$.

\begin{figure}[ht!]
    \centering
    \includegraphics[width = \textwidth]{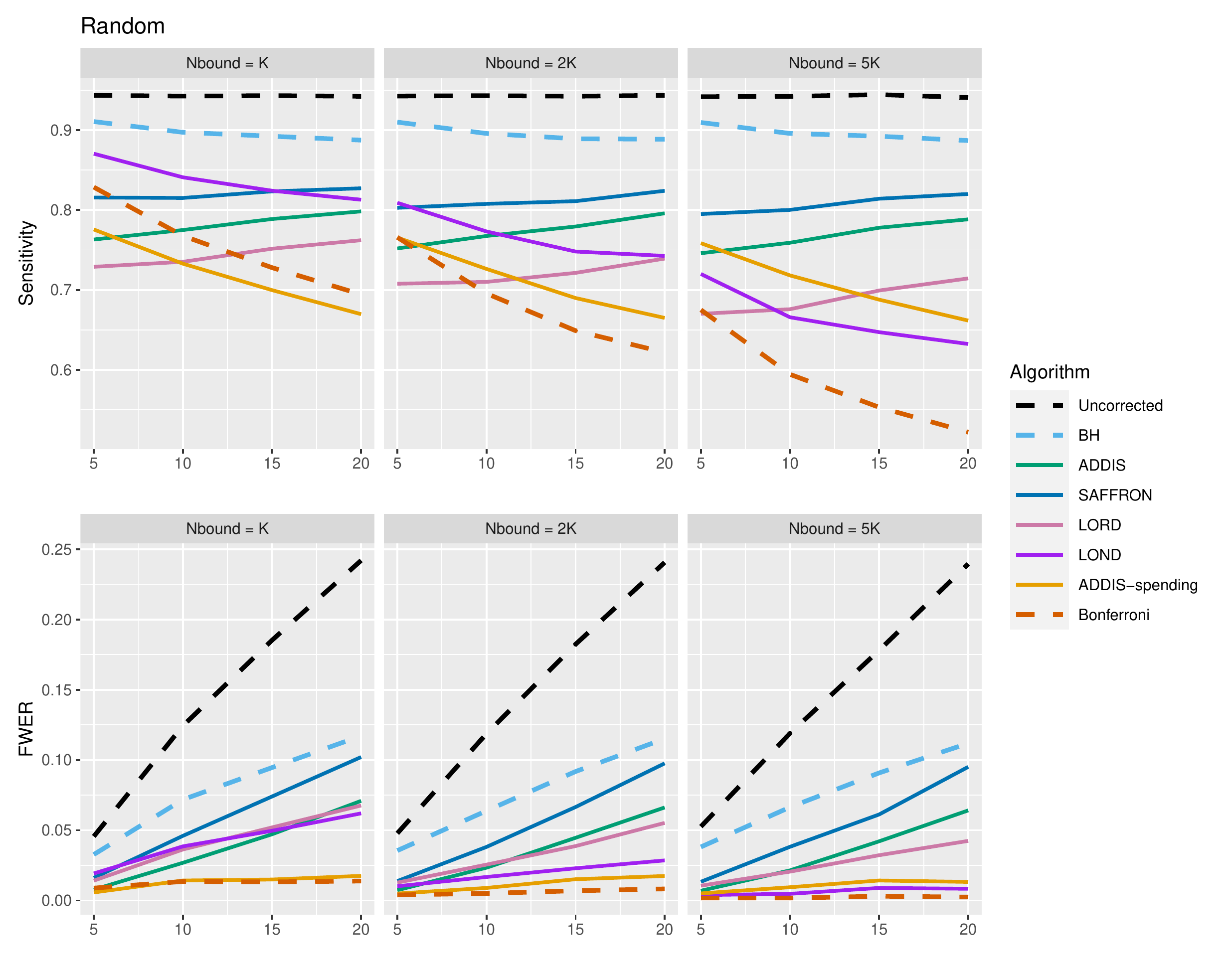}
    \caption{Sensitivity with corresponding FWER for random fixed means and ($2K/5+1$) effective treatments, with varying $\Nbound$.}
    \label{fig:fully_seq_sens_Nbound2}
\end{figure}

In terms of the effect of $\Nbound$ on the FWER, Figure~\ref{fig:fully_seq_sens_Nbound2} shows the trade-off between sensitivity and the FWER under a random ordering of treatment means. (The results for the Early and Late scenarios are given in Figures~B6 and~B7 in Section~B.2 of the Supporting web materials). We see that as $\Nbound$ increases, the FWER of Bonferroni, LOND and LORD decreases, whereas the FWER of the other algorithms remain virtually unchanged. Hence, when $\Nbound = 2K$, LOND controls the FWER at 2.5\% while still having substantial gains in power over Bonferroni. Meanwhile, the power gains of SAFFRON (and to a lesser extent, ADDIS) come at the cost of a large inflation of the FWER. However, we note that in order to be conservative it may still be useful to consider the FWER when $\Nbound = K$ since it represents a sort of `worst-case' scenario where the assumed upper bound on the total number of treatments is actually reached.

\begin{figure}[ht!]
    \centering
    \includegraphics[width = \textwidth]{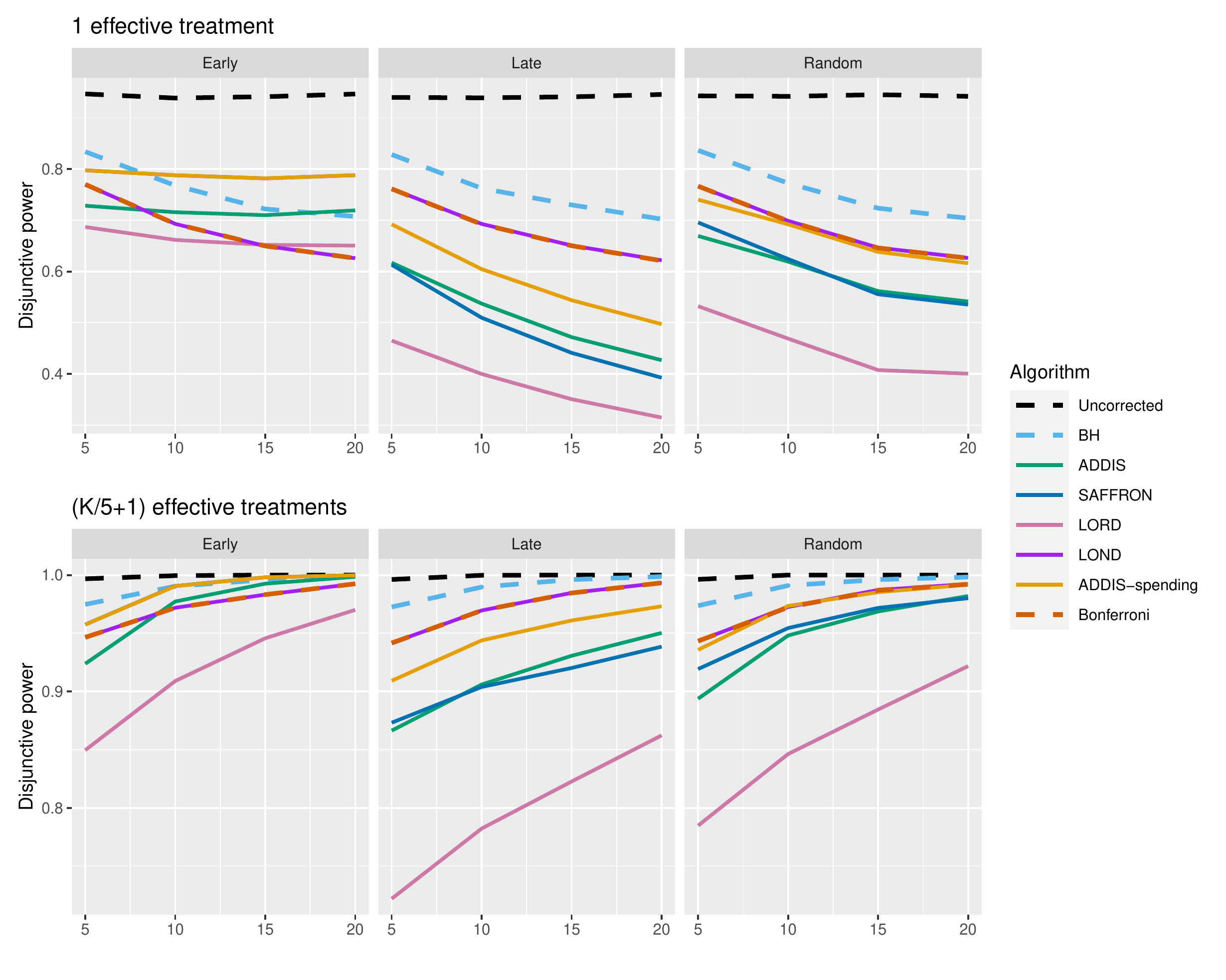}
    \caption{Disjunctive power for fixed means and different numbers of effective treatments, with $\Nbound = 2K$.}
    \label{fig:fully_seq_sens_dpower}
\end{figure}

Figure~\ref{fig:fully_seq_sens_dpower} shows the disjunctive power for~1 and $(K/5+1)$ effective treatments, with $\Nbound = 2K$ (results for $(2K/5+1)$ effective treatments are not shown since most methods have a disjunctive power very close to 1). Using this power metric, the online algorithms all have substantially lower power than Bonferroni in the Random and Late scenarios, apart from LOND (and ADDIS-spending for randomly ordered means). Only in the Early scenario do we see advantages for ADDIS-spending and (to a lesser extend) ADDIS, particularly when there is only~1 effective treatment.

\begin{figure}[ht!]
    \centering
    \includegraphics[width = \textwidth]{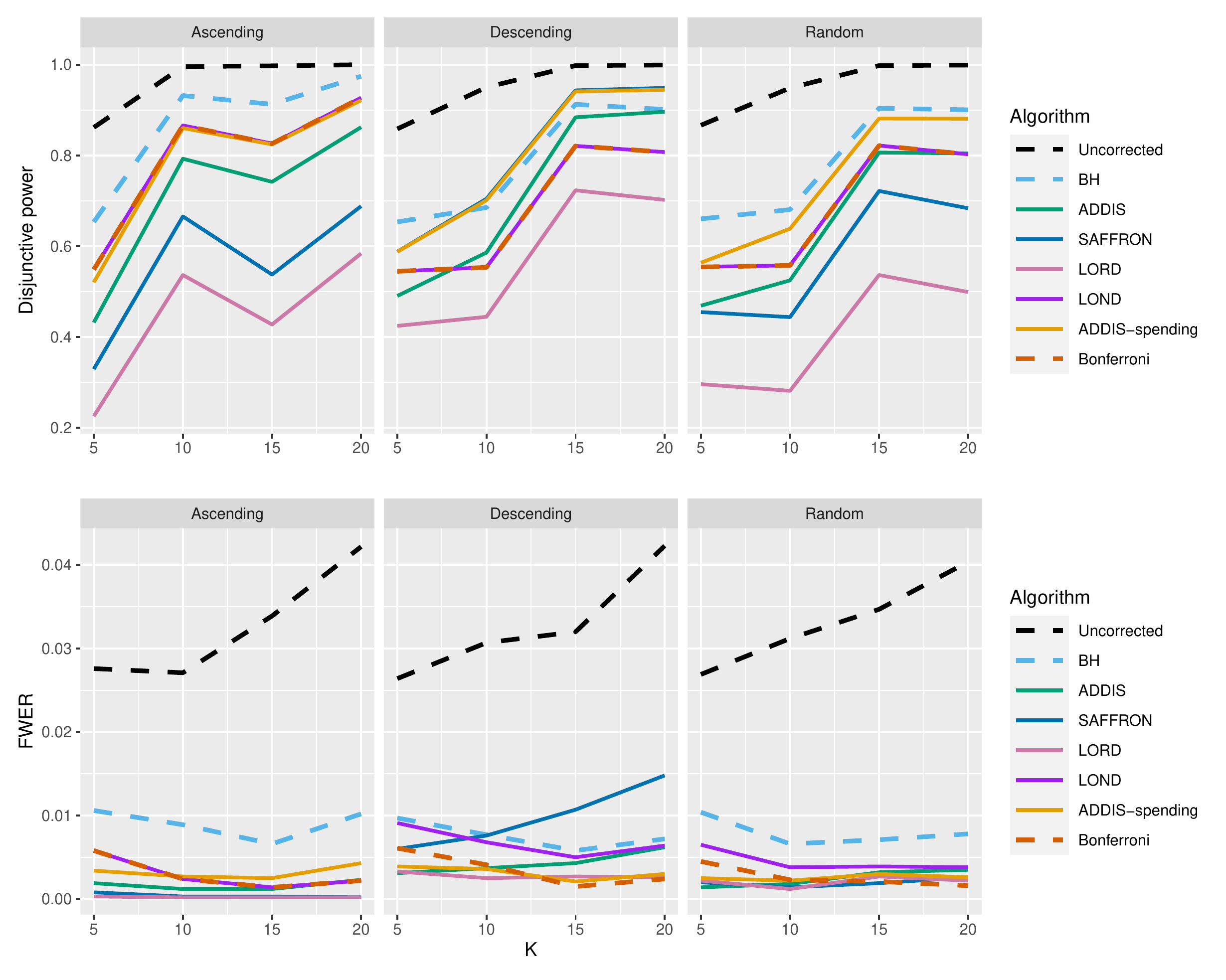}
    \caption{Disjunctive power and FWER for the staircase scenarios. Here $\Nbound = 2K$ for the disjunctive power plots and $\Nbound = K$ for the FWER plots.}
    \label{fig:staircase_dpower}
\end{figure}

Finally, we consider the impact of conservative nulls (i.e.\ the staircase scenario) in the fully sequential setting. A-priori we would expect ADDIS and ADDIS-spending to perform better here. Due to the treatment means used, the sensitivity of all algorithms is low ($<50\%$) and hence we only focus on the disjunctive power (the results for sensitivity are available in Section~B.2 of the Supporting web materials, Figure~B8). Figure~\ref{fig:staircase_dpower} shows the results for the staircase scenario, where $\Nbound = 2K$ for the disjunctive power plots and $\Nbound = K$ for the FWER plots. In this setting and for this power metric, ADDIS-spending performs particularly well with comparable disjunctive power to Bonferroni in the worst case (Ascending) and noticeable increases in power under the Random and Descending staircase scenarios. Note that the FWER of all the procedures (except Uncorrected testing) control the FWER below the nominal 2.5\% level.

In summary, we have seen that the performance of the fully sequential online testing algorithms strongly depends on the number of effective treatments, the order in which they are tested in the trial, and the assumed upper bound $\Nbound$ on the total number of treatment arms. As would be expected, the power of the online algorithms increases when there is a relatively large number of effective treatments which appear early on in the trial. However, as is the case for offline FDR controlling procedures, the online FDR algorithms do not control the FWER, although the FWER is still noticeably lower than with uncorrected testing.

Overall, LOND is a robust choice across the scenarios considered, since it provably always make at least as many rejections as Bonferroni, which can sometimes translate into substantial sensitivity gains, while maintaining a reasonable FWER for all but the most extreme cases. For example, under (arguably) the most plausible setting of a random ordering of the truly effective treatment, when $\Nbound = 2K$ even when $\Nbound = 2K$ the FWER on LOND is below the nominal 2.5\%. Even when $\Nbound = K$, the FWER is still below 5\% for $K \leq 15$. In the staircase scenario though, ADDIS-spending performs surprisingly well in terms of disjunctive power, with a noticeably higher power than LOND and Bonferroni in the Random and Descending scenarios (while still maintaining FWER control). We return to our general recommendations in the Discussion (Section~\ref{sec:discuss}).

\clearpage

\subsection{Batched algorithms}
\label{subsec:batched}

\begin{figure}[ht]
    \centering
    \includegraphics[width = \textwidth]{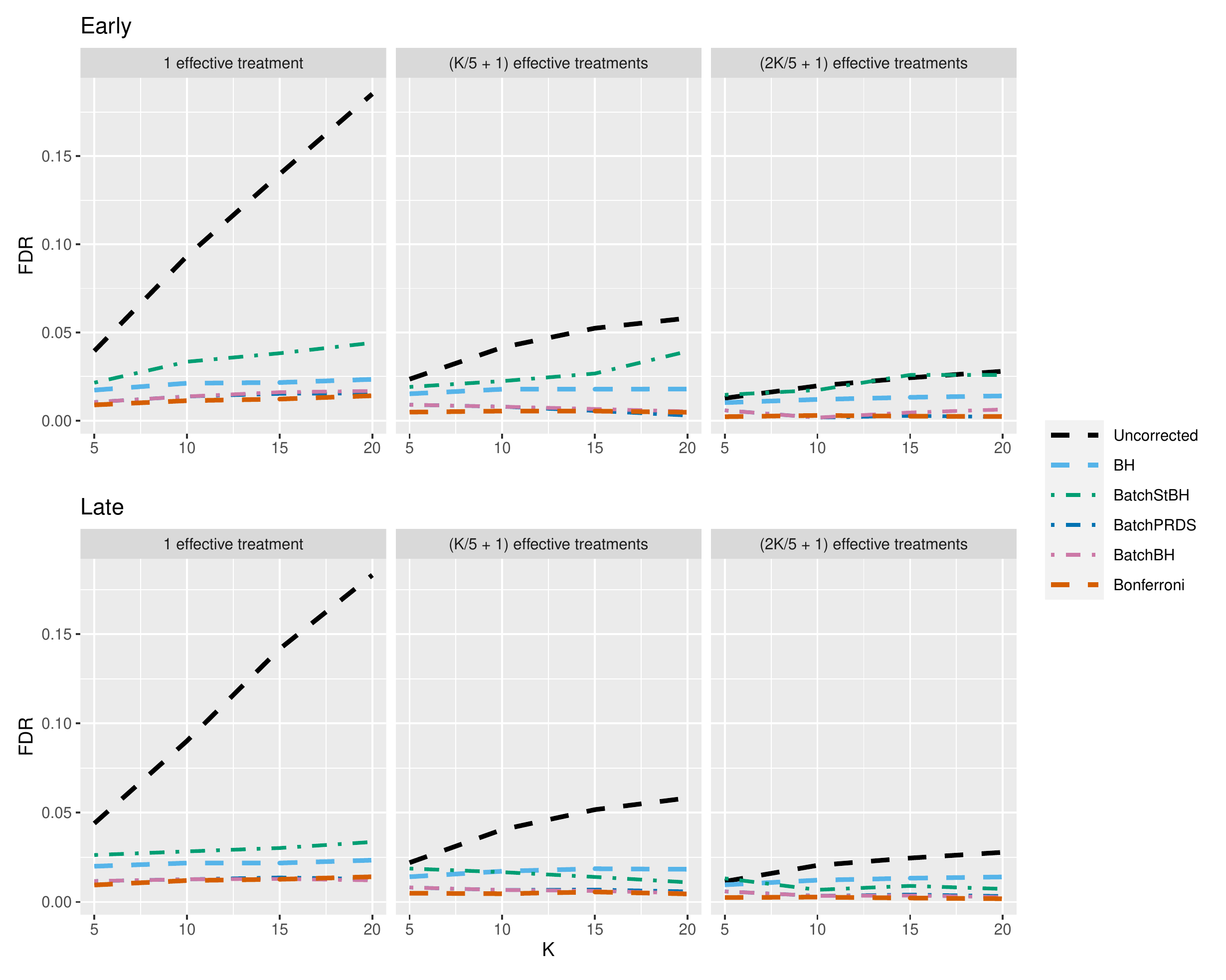}
    \caption{FDR for the batched online algorithms, with $\Nbound = K$.}
    \label{fig:batched_FDR}
\end{figure}

We now turn our attention to the batched setting (with a batch size of~5) in order to assess how the online batched algorithms (BatchBH, BatchPRDS and BatchStBH) compare with the fully sequential algorithms. Starting with the FDR, Figure~\ref{fig:batched_FDR} shows the worst and best case FDR (corresponding to the Early and Late scenarios, respectively) of the batched algorithms when $\Nbound = K$. We see that while BatchBH and BatchPRDS control the FDR below the nominal 2.5\% in all the scenarios, BatchStBH has an inflated FDR when there is 1~effective treatment which approaches 5\% in the worst case when $K = 20$. This inflation of the FDR of BatchStBH is theoretically justified, since BatchStBH only controls the FDR under independence of the $p$-values both within and across batches.

\begin{figure}[ht!]
    \centering
    \includegraphics[width = \textwidth]{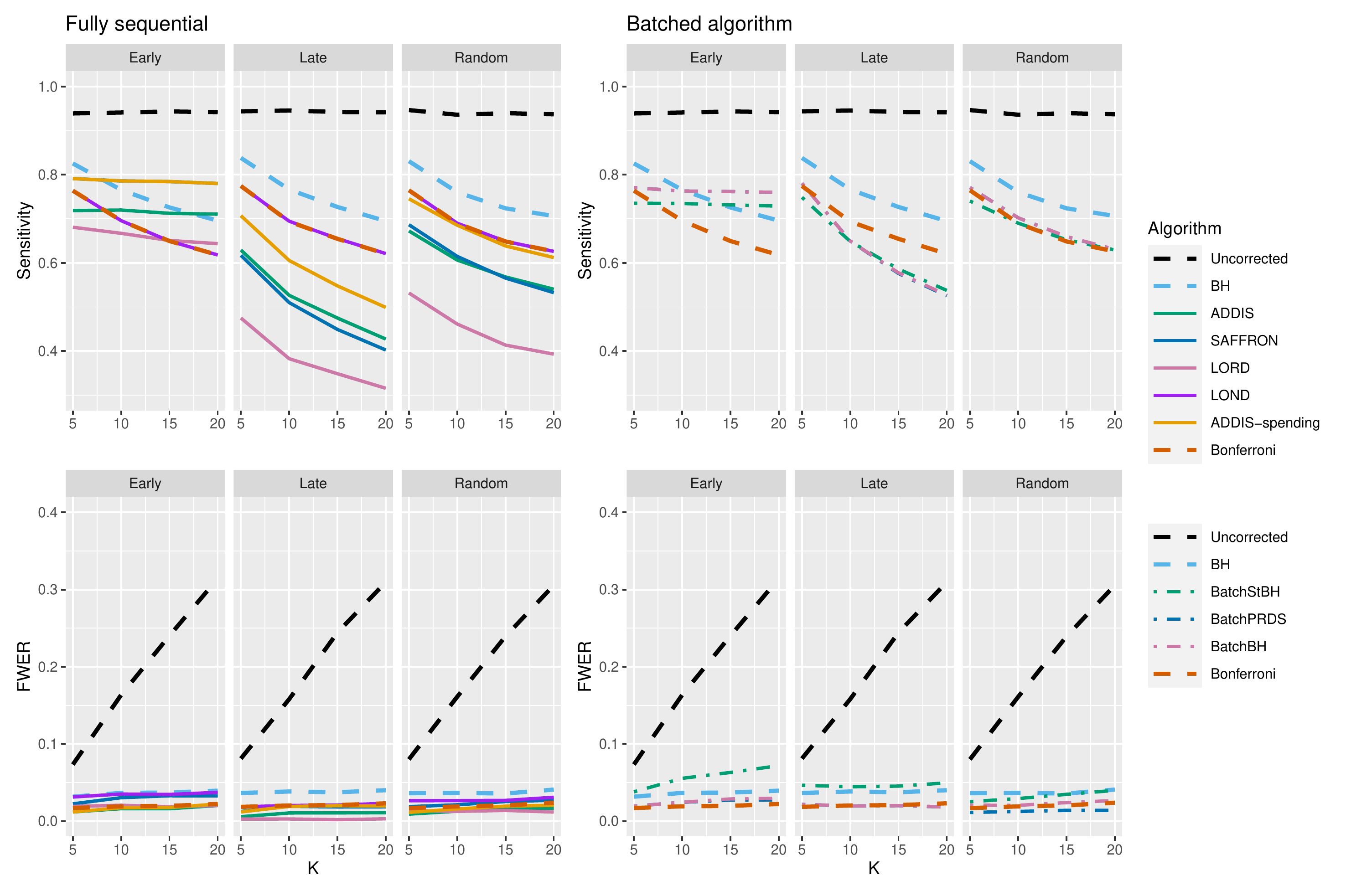}
    \caption{Comparison of sensitivity and FWER for fully sequential and batched online algorithms, with 1 effective treatment. Here $\Nbound = 2K$ for the sensitivity and $\Nbound = K$ for the FWER.}
    \label{fig:batched_sens_FWER1}
\end{figure}

\begin{figure}[ht!]
    \centering
    \includegraphics[width = \textwidth]{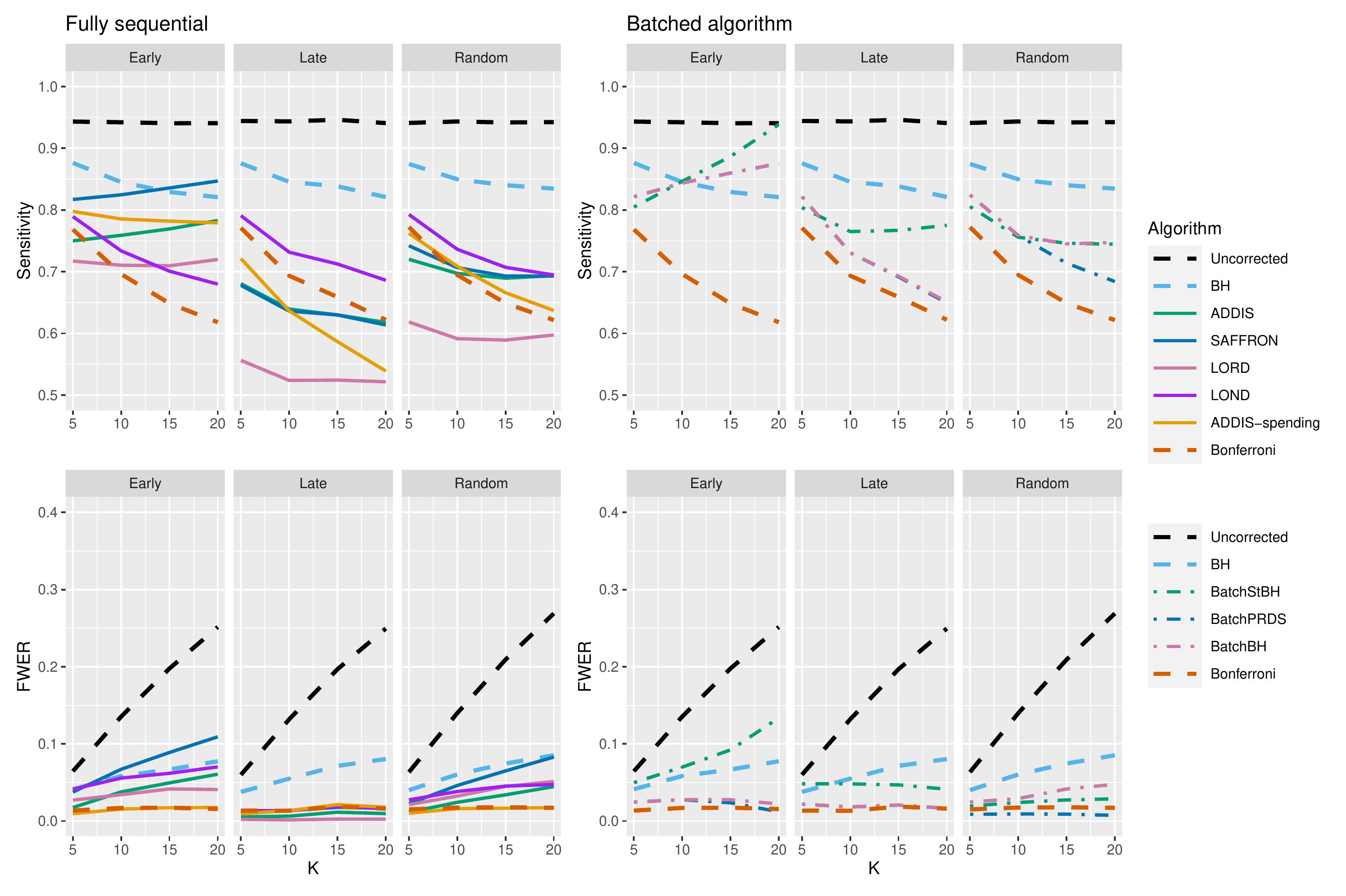}
    \caption{Comparison of sensitivity for fully sequential and batched online algorithms, with $(K/5+1)$ effective treatments. Here $\Nbound = 2K$ for the sensitivity and $\Nbound = K$ for the FWER.}
    \label{fig:batched_sens_FWER2}
\end{figure}

\begin{figure}[ht!]
    \centering
    \includegraphics[width = \textwidth]{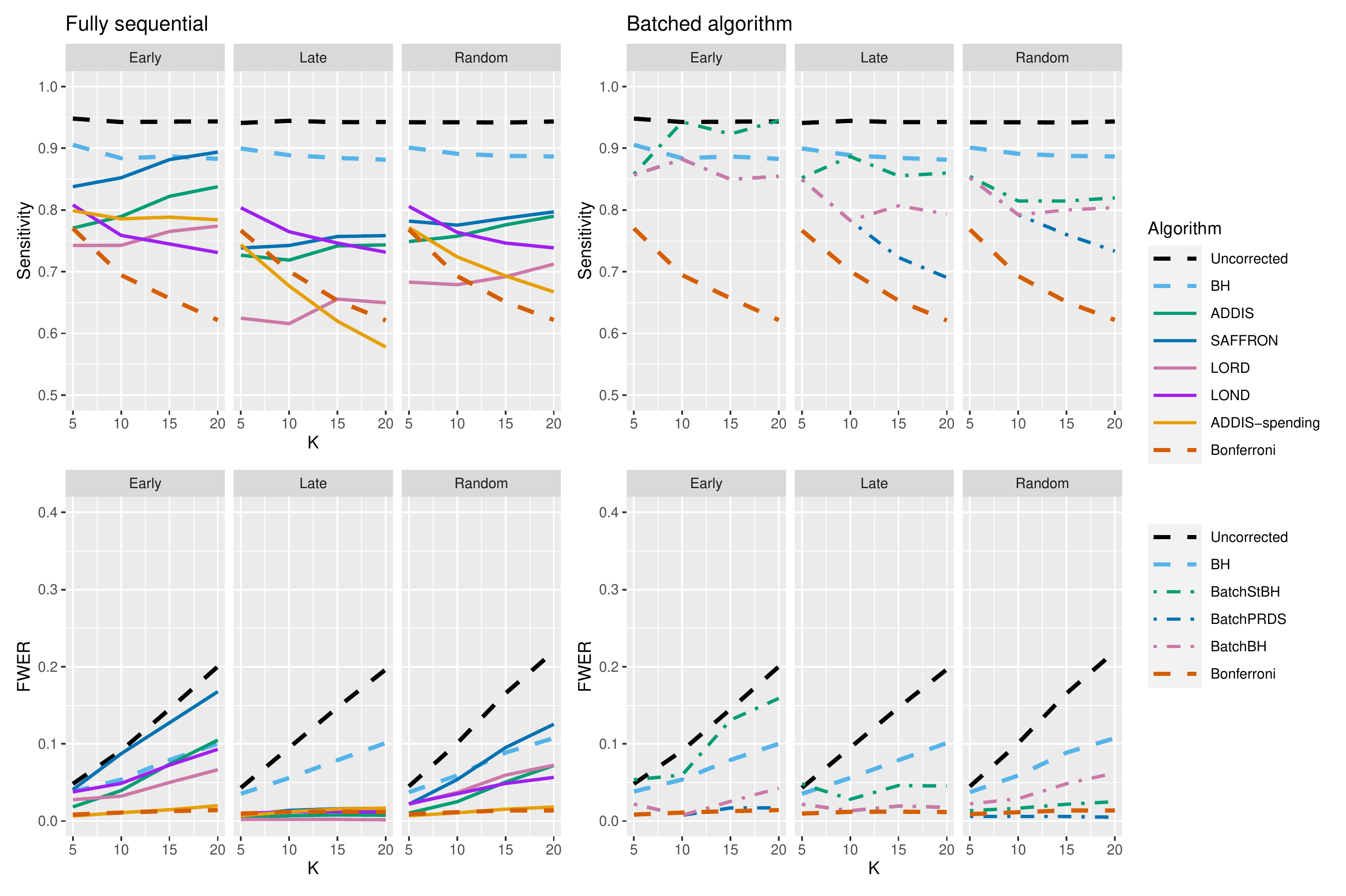}
    \caption{Comparison of sensitivity for fully sequential and batched online algorithms, with $(2K/5+1)$ effective treatments. Here $\Nbound = 2K$ for the sensitivity and $\Nbound = K$ for the FWER.}
    \label{fig:batched_sens_FWER3}
\end{figure}

Figures~\ref{fig:batched_sens_FWER1}--\ref{fig:batched_sens_FWER3} compares the sensitivity and FWER for fully sequential and batched online algorithms, with $\Nbound = 2K$ for the sensitivity and $\Nbound = K$ for the FWER. Starting first with 1~effective treatment (Figure~\ref{fig:batched_sens_FWER1}), there is no advantage in terms of sensitivity when using the batched algorithms compared to LOND under the Late and Random scenarios, or ADDIS-spending and SAFFRON in the early scenario. In all scenarios, BatchStBH has a noticeably inflated FWER, particularly under the Early scenario. In contrast, BatchBH and BatchPRDS are reasonably close to the nominal 2.5\% level.


When there are $(K/5+1)$ effective treatments (Figure~\ref{fig:batched_sens_FWER2}), there is a slight sensitivity advantage in using BatchBH and BatchStBH over LOND in the Random scenario for $K > 10$, with similar or lower FWER. In the Early scenario, BatchBH and BatchPRDS have similar sensitivity to SAFFRON but substantially lower FWER (of approximately 2.5\%). Meanwhile, BatchStBH has a higher sensitivity than SAFFRON for $K>10$, but at the cost of an even more inflated FWER. Finally, in the Late scenario, BatchBH and BatchPRDS have a slightly lower sensitivity than SAFFRON, whereas BatchStBH has a substantially higher sensitivity but at the cost of a noticeable inflation of the FWER.

When there are $(2K/5+1)$ effective treatments (Figure~\ref{fig:batched_sens_FWER3}), BatchPRDS has similar sensitivity to LOND but with a substantially lower FWER that is below the 2.5\% level. BatchBH and BatchStBH have similar sensitivity to SAFFRON and ADDIS respectively when $K \geq 10$, but BatchStBH has substantially lower FWER (approximately 2.5\%) while BatchBH has similar FWER to ADDIS. In the Early scenario, BatchBH and BatchPRDS have a roughly similar sensitivity to SAFFRON, but with a substantially lower FWER (which as at or below 2.5\% for BatchPRDS). BatchStBH has a higher sensitivity than SAFFRON (in fact approaching the level of uncorrected testing) but with a similarly highly inflated FWER. Lastly, in the Late scenario BatchPRDS has a lower sensitivity than LOND for $K \geq 10$ although still noticeably higher than Bonferroni. BatchBH has a slightly higher sensitivity than LOND with a similar FWER. Meanwhile BatchStBH has a substantially higher sensitivity than LOND (or any of the other fully sequential algorithms) but again at the cost of a noticeably higher FWER.

In terms of disjunctive power, Figure~B9 in Section~B.3 of the Supporting web materials shows that all of the batched algorithms have fairly similar power to ADDIS-spending across the different scenarios. However, as already seen this comes at the cost of a (potentially highly) inflated FWER for BatchStBH and to a lesser extent BatchBH.

\begin{sidewaysfigure}[ht]
    \centering
    \includegraphics[width = \textwidth]{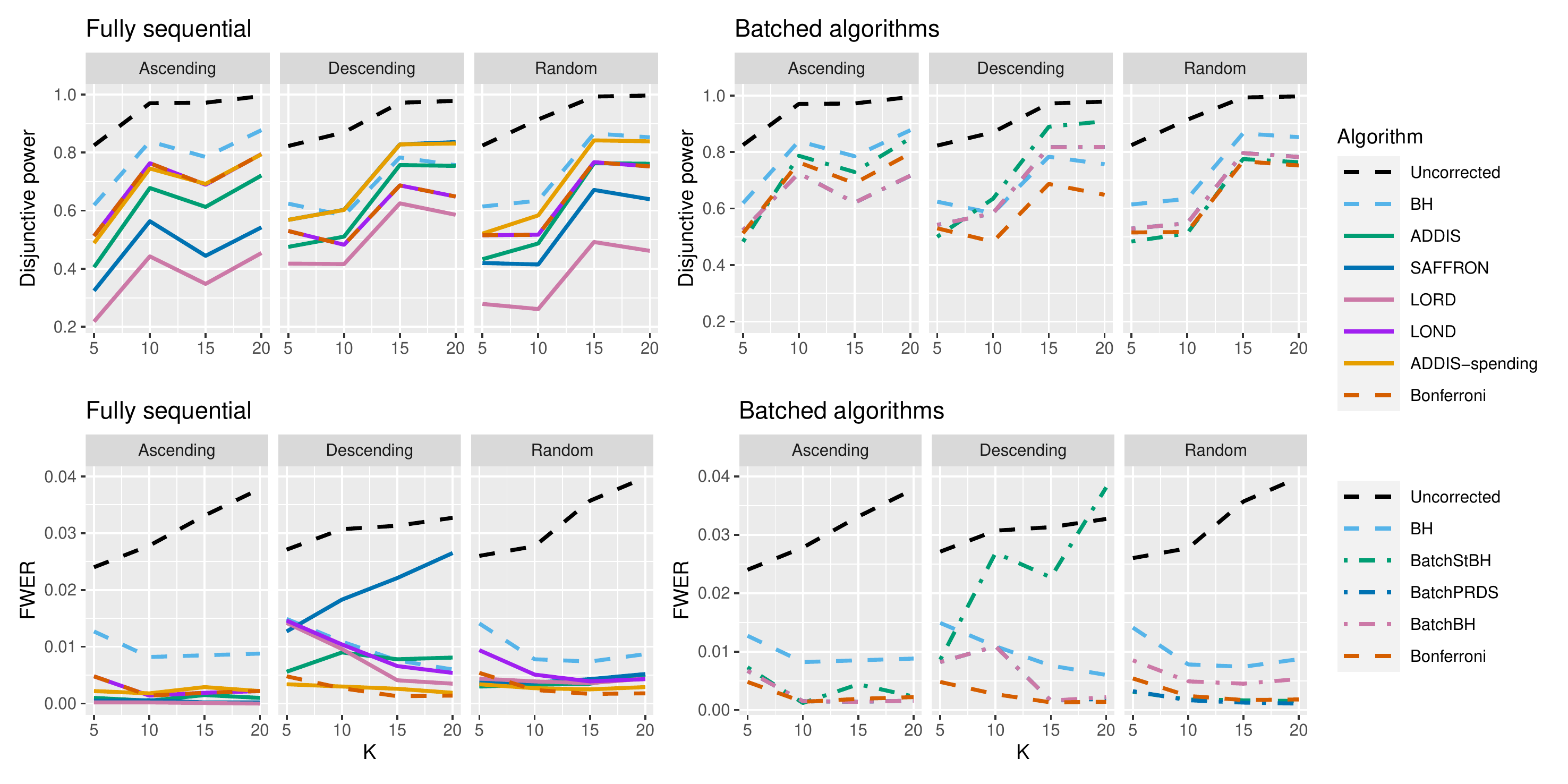}
    \caption{Comparison of the FWER ($\Nbound = K$) and disjunctive power ($\Nbound = 2K$) for fully sequential and batched online algorithms under the staircase scenario.}
    \label{fig:batched_staircase_dpower}
\end{sidewaysfigure}

Finally, Figure~\ref{fig:batched_staircase_dpower} gives the results for the staircase scenario with conservative nulls, in terms of disjunctive power and FWER. We again see that the Batched algorithms tend to have similar disjunctive power to ADDIS-spending, although BatchStBH has slightly higher values in the Ascending and Descending scenarios and all the batched algorithms have slightly lower values in the Random scenarios. As for the FWER, all the batched algorithms control the FWER below the nominal 2.5\%, except for BatchStBH under the Descending scenario.

Looking at the results for batched testing as a whole, we see that the BatchPRDS algorithm is competitive when compared to LOND (and hence also Bonferroni) by having similar or higher power, in terms of sensitivity and disjunctive power, while still controlling the FWER at 2.5\% cross all the scenarios considered.

\clearpage

\section{Case study: STAMPEDE trial}
\label{sec:case_study}

\begin{figure}[ht!]
    \centering
    \includegraphics[width = \textwidth]{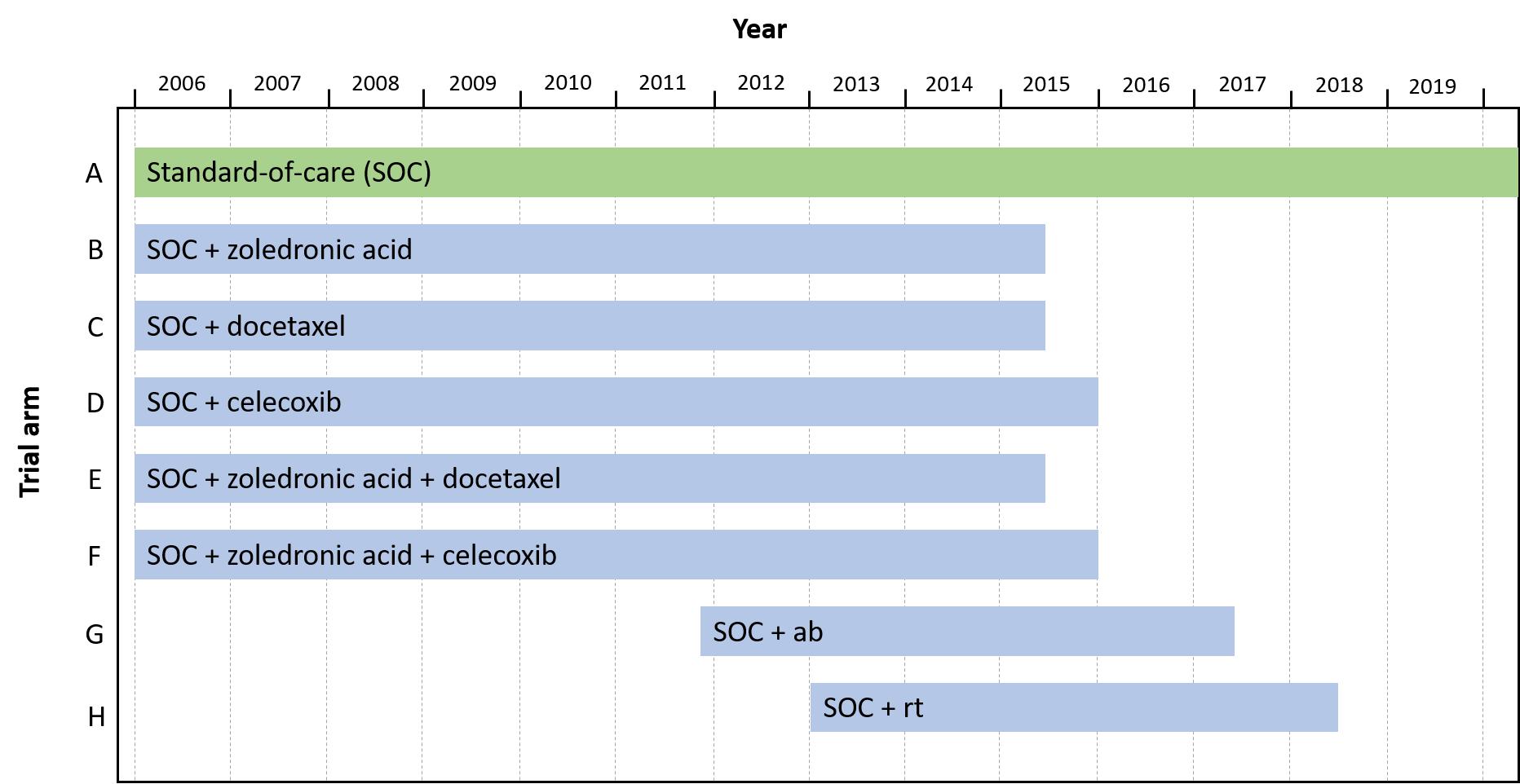}
    \caption{Schematic of the STAMPEDE trial. ab = abiraterone, rt = radiotherapy.}
    \label{fig:STAMPEDE}
\end{figure}

The STAMPEDE (Systemic Therapy for Advancing or Metastatic Prostate Cancer) trial is a flagship platform trial for research into the effect of systemic therapies for prostate cancer~\cite{james2008stampede} on overall survival. The trial opened to accrual in October 2005 with 6 arms (A--F), including the control arm~A, which was standard-of-care (SOC) hormone therapy. Figure~\ref{fig:STAMPEDE} shows a schematic of the treatment comparisons that have already been reported from STAMPEDE. Two additional arms (G and H) were added to the trial in 2011 and 2013, respectively. Three arms (B, C and E) reported main analyses in 2015~\cite{james2016stampede}, with two additional arms (D and F) reporting in 2015/16~\cite{mason2017stampede}. Arm~G reported main analyses in 2017~\cite{james2017stampede} and Arm~H reported results in 2018~\cite{parker2018stampede}.

Table~\ref{tab:STAMPEDE_pval} shows the reported $p$-values (unadjusted for multiplicity) when comparing arms~B to~H with the SOC (arm~A), as given in \cite{james2016stampede, mason2017stampede, james2017stampede, parker2018stampede}. The dashed lines denote the 4 batches in the trial. 

\begin{table}[ht!]
    \centering
    \begin{tabular}{l l}
         \textbf{Trial arm} & \textbf{$p$-value} \\ \hline
         \textbf{B}: SOC + zoledronic acid & 0.450 \\
         \textbf{C}: SOC + docetaxel & 0.006 \\
         \textbf{E}: SOC + zoledronic acid + docetaxel & 0.022 \\ \hdashline
         \textbf{D}: SOC + celecoxib & 0.847 \\
         \textbf{F}: SOC + zoledronic acid + celecoxib & 0.130 \\ \hdashline
         \textbf{G}: SOC + abiraterone & 0.001 \\ \hdashline
         \textbf{H}: SOC + radiotherapy & 0.266 \\
    \end{tabular}
    \caption{Reported results for the STAMPEDE trial. SOC = Standard-of-care.}
    \label{tab:STAMPEDE_pval}
\end{table}

We now apply the offline and online testing algorithms to the observed $p$-values given above in Table~\ref{tab:STAMPEDE_pval}, keeping the alphabetical ordering of $p$-values within the three batches. As a sensitivity analysis, we show what happens if the ordering of the $p$-values in the first batch changes in Table~C2 in Section~C of the Supporting web materials. We set the upper bound on the number of treatments $\Nbound = 20$, i.e.\ twice as many arms that have already entered the STAMPEDE trial as of the end of 2021
.

Table~\ref{tab:STAMPEDE_rej} shows which of the hypotheses corresponding to each trial arm can be rejected at level $\alpha \in \{0.025, 0.05, 0.1\}$, as well as the current significance level $\alpha_8$ that would be used to test the next treatment arm after the 7 already evaluated in the trial. We consider larger values of $\alpha$ since this has been a suggestion for multi-arm trials~\cite{wason2016} as a compromise between not correcting for multiplicity at all (as would be the case when running a series completely independent two-arm trials) and strict FWER control at 2.5\%.
When $\alpha = 0.025$, uncorrected testing rejects hypotheses C, E and~G, but only BH and BatchStBH reject more than one hypothesis (C and G). Bonferroni and all the other online algorithms only reject hypothesis~G, except for ADDIS and LORD which reject no hypotheses. When $\alpha = 0.05$ the rejections made by the various algorithms remain the same, except now SAFFRON, BatchBH and BatchPRDS reject hypotheses~C and~G, while BatchStBH rejects hypotheses C, E and~G. Finally, when $\alpha = 0.1$ we see that BH, SAFFRON, BatchBH, BatchPRDS and BatchStBH all make the same rejections as Uncorrected testing. In terms of the current adjusted testing level $\alpha_8$, we see that ADDIS-spending, ADDIS and LORD have a smaller value than even Bonferroni. SAFFRON, LOND, BatchBH and BatchPRDS have a larger value of $\alpha_8$ than Bonferroni, but only BatchStBH has a higher value than Uncorrected testing.\\

\begin{table}[ht!]
    \centering
    \begin{tabular}{l | l l l | l l l}
         \textbf{Algorithm} & \multicolumn{3}{c}{\textbf{Hypotheses rejected}} & \multicolumn{3}{|c}{$\bm{\alpha_8}$} \\
         & $\alpha = 0.025$ & $\alpha = 0.05$ & $\alpha = 0.1$ & $\alpha = 0.025$ & $\alpha = 0.05$ & $\alpha = 0.1$ \\ \hline
         Uncorrected & C, E, G & C, E, G & C, E, G & 0.0250 & 0.0500 & 0.1000\\
         Bonferroni & G & G & G & 0.0013 & 0.0025 & 0.0050 \\
         ADDIS-spending & G & G & G & 0.0005 & 0.0011 & 0.0021\\
         BH & C, G & C, G & C, E, G & -- & -- & -- \\
         ADDIS & -- & G & G & 0.0003 & 0.0016 & 0.0031\\
         SAFFRON & G & C, G & C, E, G & 0.0041 & 0.0165 & 0.0412\\
         LORD & -- & -- & -- & 0.0001 & 0.0002 & 0.0003\\
         LOND & G & G & G & 0.0025 & 0.0050 & 0.0100\\
         BatchBH & G & C, G & C, E, G & 0.0019 & 0.0057 & 0.0151\\
         BatchPRDS & G & C, G & C, E, G & 0.0019 & 0.0057 & 0.0151\\
         BatchStBH & C, G & C, E, G & C, E, G & 0.0381 & 0.1015 & 0.1238\\
    \end{tabular}
    \caption{Rejections and current significance level $\alpha_8$ of different algorithms using the results of the STAMPEDE trial, with the ordering as in Table~\ref{tab:STAMPEDE_pval}.}
    \label{tab:STAMPEDE_rej}
\end{table}



\section{Discussion}
\label{sec:discuss}

In this paper, we have shown how online hypothesis testing can be applied to the platform trial setting to achieve overall control of type~I errors. In many of the simulation scenarios, there were noticeable gains in sensitivity compared with using Bonferroni, although this has to be considered carefully with respect to the potential inflation of the FWER and how acceptable that may be to stakeholders (particularly from a regulatory viewpoint). However, in all cases the FWER was lower than Uncorrected testing, and so from that perspective, any of the online algorithms would offer an improvement.

We have focused on trial settings that ultimately test $N \geq 20$ hypotheses, which is a relatively small number of hypotheses compared with most previous simulation results in the literature. Our results show that the online testing framework offers a smaller advantage in terms of power properties in this small~$N$ setting compared to when a very large number ($\geq 1000$) of hypotheses are tested. However, in some clinical trial settings (e.g.\ testing for interaction with genomic data) there can be a large number of hypotheses and a desire for FDR control (as opposed to FWER control), and so online error rate control may be especially applicable.

As noted in Section~\ref{sec:simul_results}, the relative performance of the online algorithms compared with Bonferroni and Uncorrected testing crucially depends on the number of effective treatments, the order in which they are tested in the trial, as well as the assumed upper bound $\Nbound$. Hence, the `best' online algorithm to use may be quite different depending on the trial context and goals. In general, across the simulation scenarios for the fully sequential setting, we see that LOND seems to offer quite a good compromise by guaranteeing at least as many rejections as Bonferroni and often seeing noticeable increases in sensitivity, while still maintaining (depending on the trial context) reasonable FWER levels. In the batched setting however, BatchPRDS seems to be preferable to LOND as it has a similar or higher sensitivity and disjunctive power, while maintaining FWER control below the nominal level.

As seen in the case study results, in practice the online testing algorithms may not be `competitive' in terms of the number of rejections when compared with Uncorrected testing unless a more relaxed $\alpha$ is used. However, the potential inflation on top of that relaxation for many of the online algorithms also needs to be taken into account. In that sense, BatchPRDS is particularly appealing due to the lack of further FWER inflation.

We have started with the simplifying assumption that each experimental treatment is tested exactly once, as soon as the outcomes from a pre-specified number of patients have been observed. A useful extension would be to allow multiple looks (stages) for each treatment arm, with early stopping for futility and efficacy. Indeed, the STAMPEDE trial included interim analyses that we did not take into account in our analysis in Section~\ref{sec:case_study}.A complicating factor is that to directly apply the online testing algorithms, we have only assigned the adjusted testing level $\alpha_i$ to hypothesis $H_i$ at the point of testing. To allow repeated testing of hypothesis~$H_i$ and early stopping, we would require $H_i$ to be determined in advance. For further discussion and proposals for this setting, we refer the interested reader to Zrnic et al.~\cite{Zrnic2018}, who proposed a framework for asynchronous online testing. As well, very recent work by Zehetmayer et al.~\cite{zehet2021a} shows how to use LOND specifically for group-sequential platform trials.
    
    

Another assumption that we have made is that $\Nbound \geq K$, i.e.\ that we never test more treatment arms than originally planned for. In the case where $\Nbound < K$, then a simple method of accommodating this change has been proposed by~\cite[pg.~10]{Robertson2018}, but the power properties have not been explored in the context of platform trials.
Finally, we could also make further comparisons of online testing with existing methodologies for controlling the FWER when adding treatment arms to a platform trial~\cite{Burnett2020, ChoodariOskooei2020}.

\section*{Acknowledgements}
This research was supported by the NIHR Cambridge Biomedical Research Centre (BRC1215-20014). The views expressed in this publication are those of the authors and not necessarily those of the NHS, the National Institute for Health Research or the Department of Health and Social Care (DHCS). David S.~Robertson and Thomas Jaki received funding from the UK Medical Research Council (MC\_UU\_00002/14). David S.~Robertson also received funding from the Biometrika Trust. 

Franz König and Martin Posch were supported by EU-PEARL. EU-PEARL (EU Patient-cEntric clinicAl tRial pLatforms) project has received funding from the Innovative Medicines Initiative (IMI) 2 Joint Undertaking (JU) under grant agreement No 853966. This Joint Undertaking receives support from the European Union’s Horizon 2020 research and innovation programme and EFPIA and Children’s Tumor Foundation, Global Alliance for TB Drug Development non-profit organisation, Springworks Therapeutics Inc. This publication reflects the authors’ views. Neither IMI nor the European Union, EFPIA, or any Associated Partners are responsible for any use that may be made of the information contained herein.

\section*{Data availability statement}
All of the data that support the findings of this study are available within the paper itself.



\printbibliography

\end{document}


\maketitle

\renewcommand{\thesection}{\Alph{section}}
\renewcommand\thefigure{\thesection\arabic{figure}}
\renewcommand\thetable{\thesection\arabic{table}}    
\setcounter{table}{0}   
\setcounter{figure}{0}  

\section{Online testing algorithms implementation}
\label{appendix:implementation}

For all the simulations, we use the \texttt{onlineFDR} R package in order to implement the online testing algorithms. Table~\ref{tab:online_implementation} below gives the parameter values used for each of the algorithms. All of the $\gamma_i$ sequences are chosen so that $\displaystyle \sum_{i=1}^{\Nbound} \gamma_i = 1$.

\begin{table}[ht!]
    \centering
    \renewcommand{\arraystretch}{1.5}
    \begin{tabular}{l l l }
         \textbf{Algorithm} &  $\gamma_i$ \textbf{sequence} & \textbf{Other parameters} \\[6pt] \hline
         ADDIS-spending & $\gamma_i \propto 1/i^{1.6}$ & $\lambda = 0.25$, $\tau = 0.5$\\
         ADDIS & $\gamma_i \propto 1/i^{1.6}$ & $\lambda = \tau = 0.5$, $w_0 = \lambda \tau \alpha/2$\\
         LOND & $\gamma_i \propto 1$ \\
         LORD & $\gamma_i \propto \frac{\log(\max(i,2))}{i \, \exp(\sqrt{\log(i)})}$ & $w_0 = \alpha/10$, $b_0 = \alpha - w_0$\\[6pt]
         SAFFRON & $\gamma_i \propto 1/i^{1.6}$ & $\lambda = 0.5$, $w_0 = \alpha/2$\\
         BatchBH & $\gamma_i \propto 1/i^{1.6}$ &\\
         BatchPRDS & $\gamma_i \propto 1/i^{1.6}$ & \\
         BatchStBH & $\gamma_i \propto 1/i^{1.6}$ & $\lambda = 0.5$ \\ \hline
    \end{tabular}
    \caption{Parameter values used for the online testing algorithms.}
    \label{tab:online_implementation}
\end{table}

\newpage 
\section{Additional simulation results}

\subsection{Global null}
\label{appendix:global_null}

\begin{figure}[ht]
    \centering
    \includegraphics[width = \textwidth]{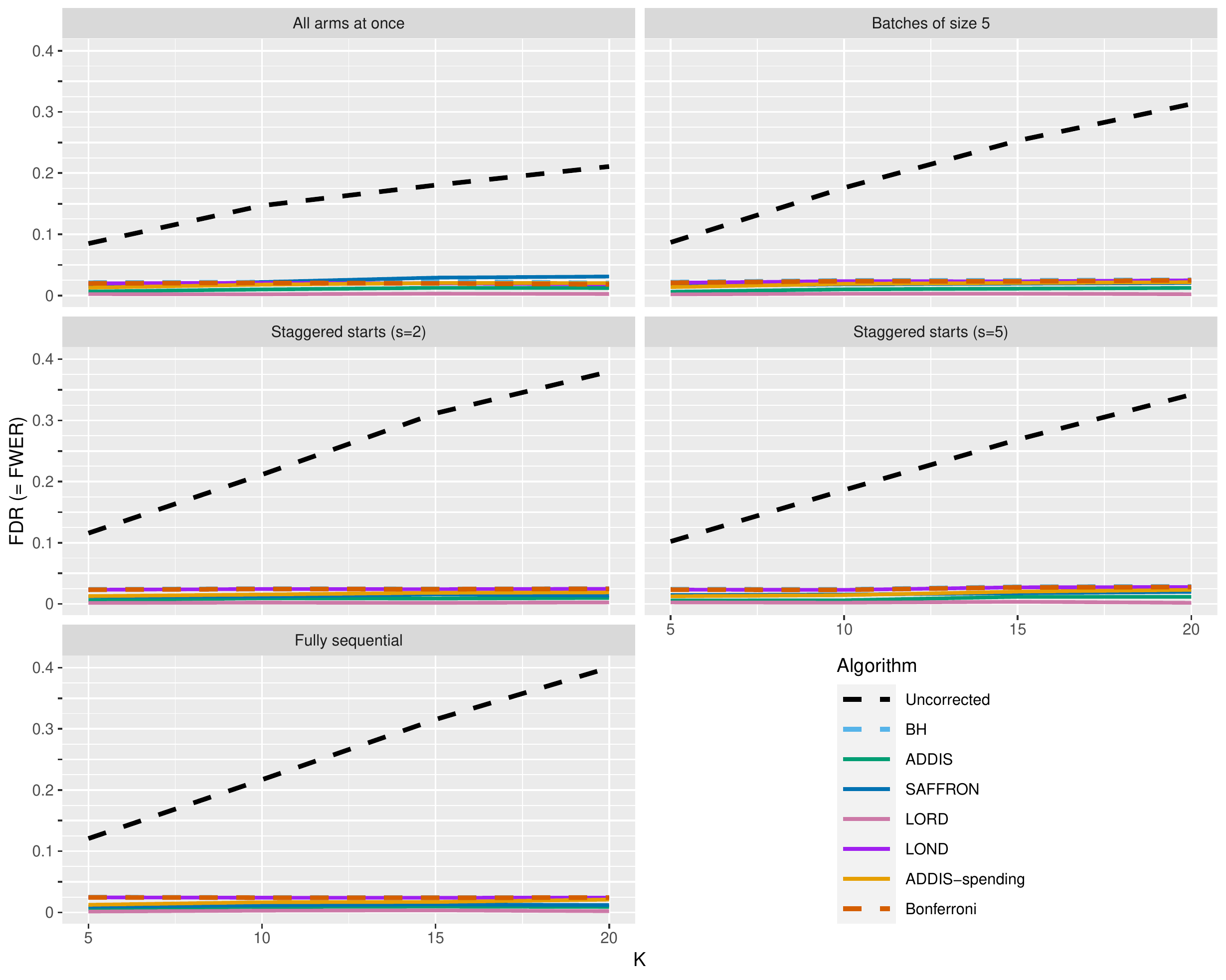}
     \caption{FDR (= FWER) under the global null with varying patterns of arm entry times and $\Nbound = K$.}
    \label{fig:global_null}
\end{figure}

\newpage

\subsection{Fully sequential setting}
\label{appendix:fully_seq}

\begin{figure}[ht]
    \centering
    \includegraphics[width = \textwidth]{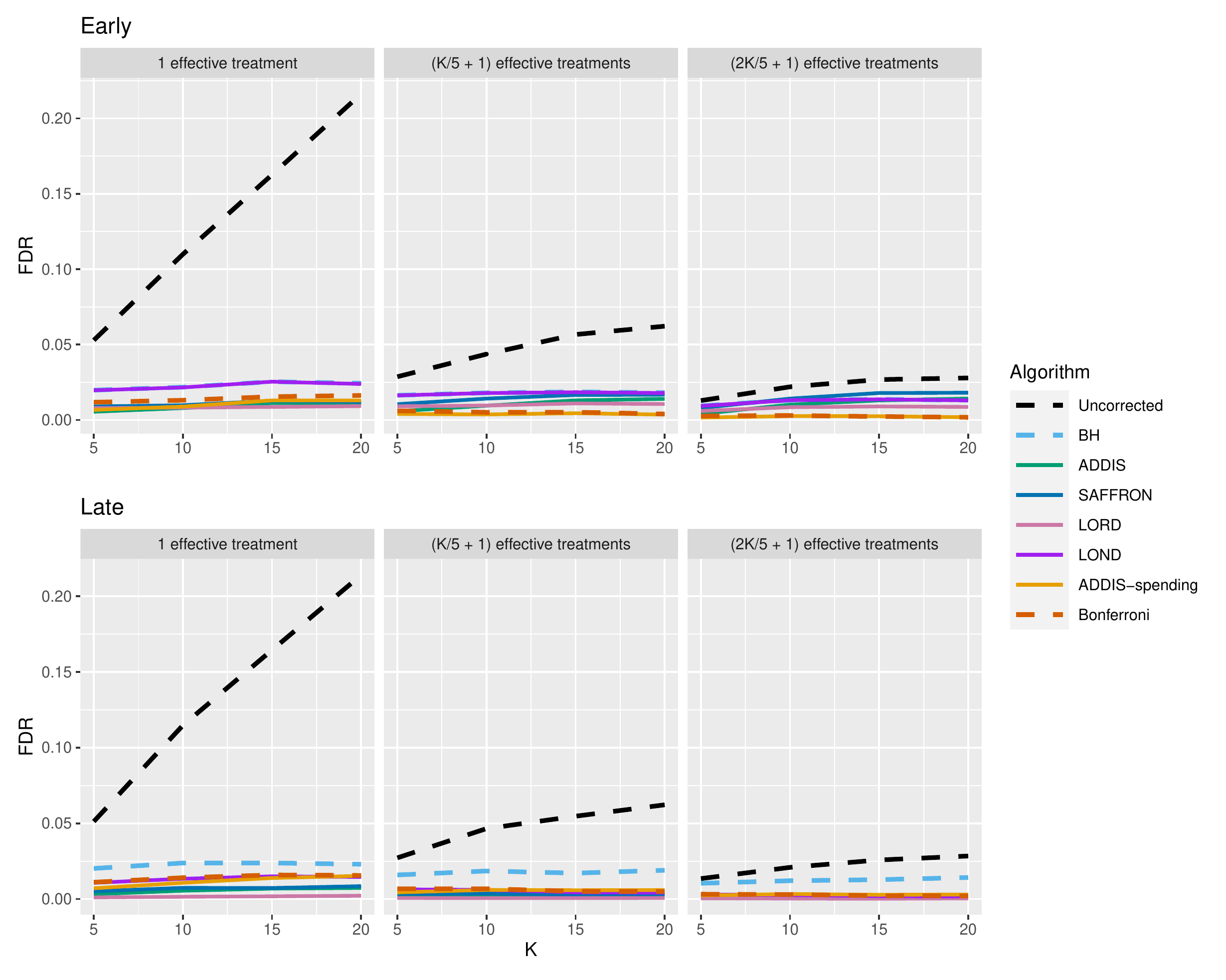}
    \caption{FDR for fixed means and different numbers of effective treatments, with $\Nbound = K$.}
    \label{fig:fully_seq_FDR}
\end{figure}

\begin{figure}[ht]
    \centering
    \includegraphics[width = \textwidth]{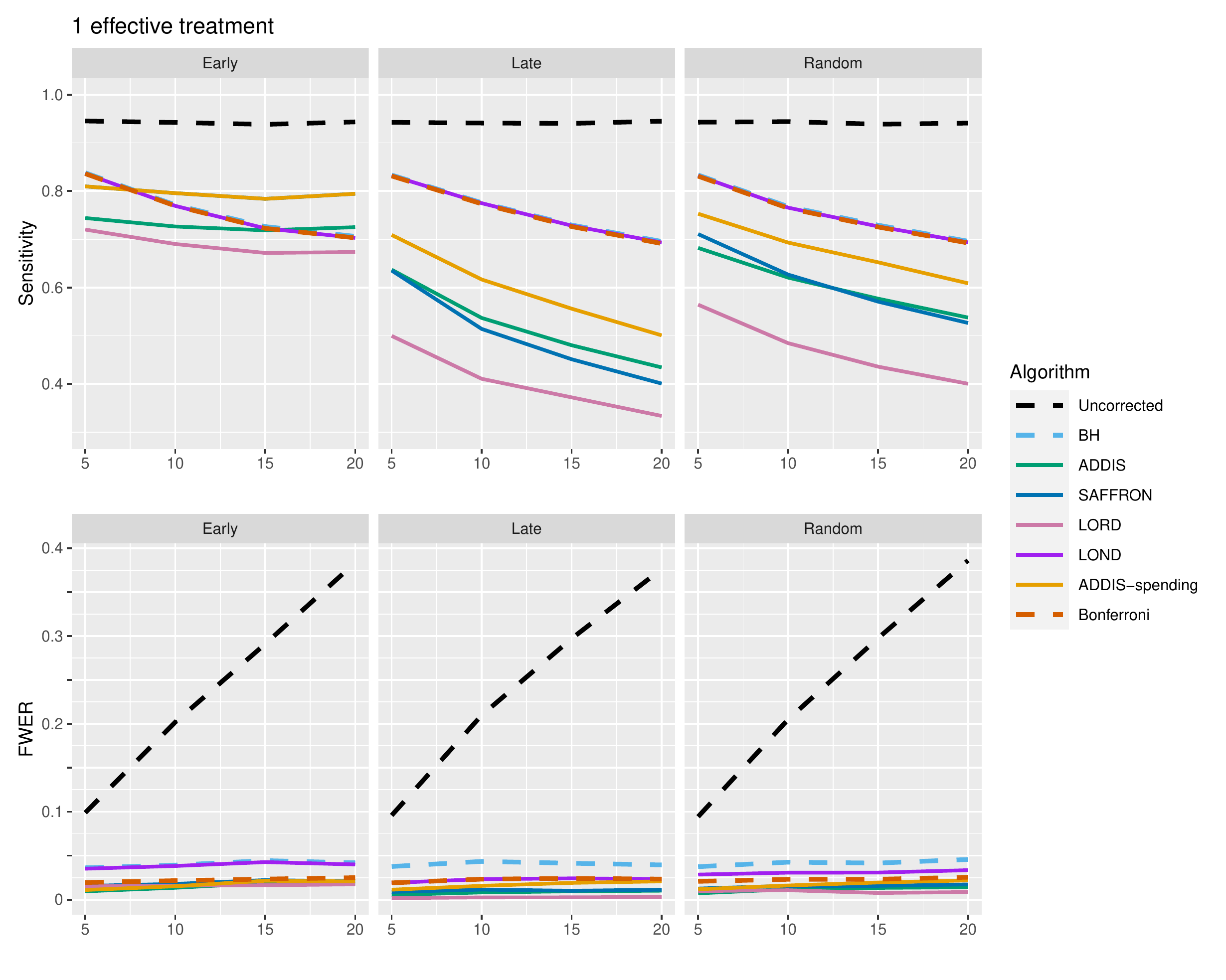}
    \caption{Sensitivity and FWER for fixed means and 1 effective treatments, with $\Nbound = K$.}
    \label{fig:fully_seq_sens_FWER1}
\end{figure}

\begin{figure}[ht]
    \centering
    \includegraphics[width = \textwidth]{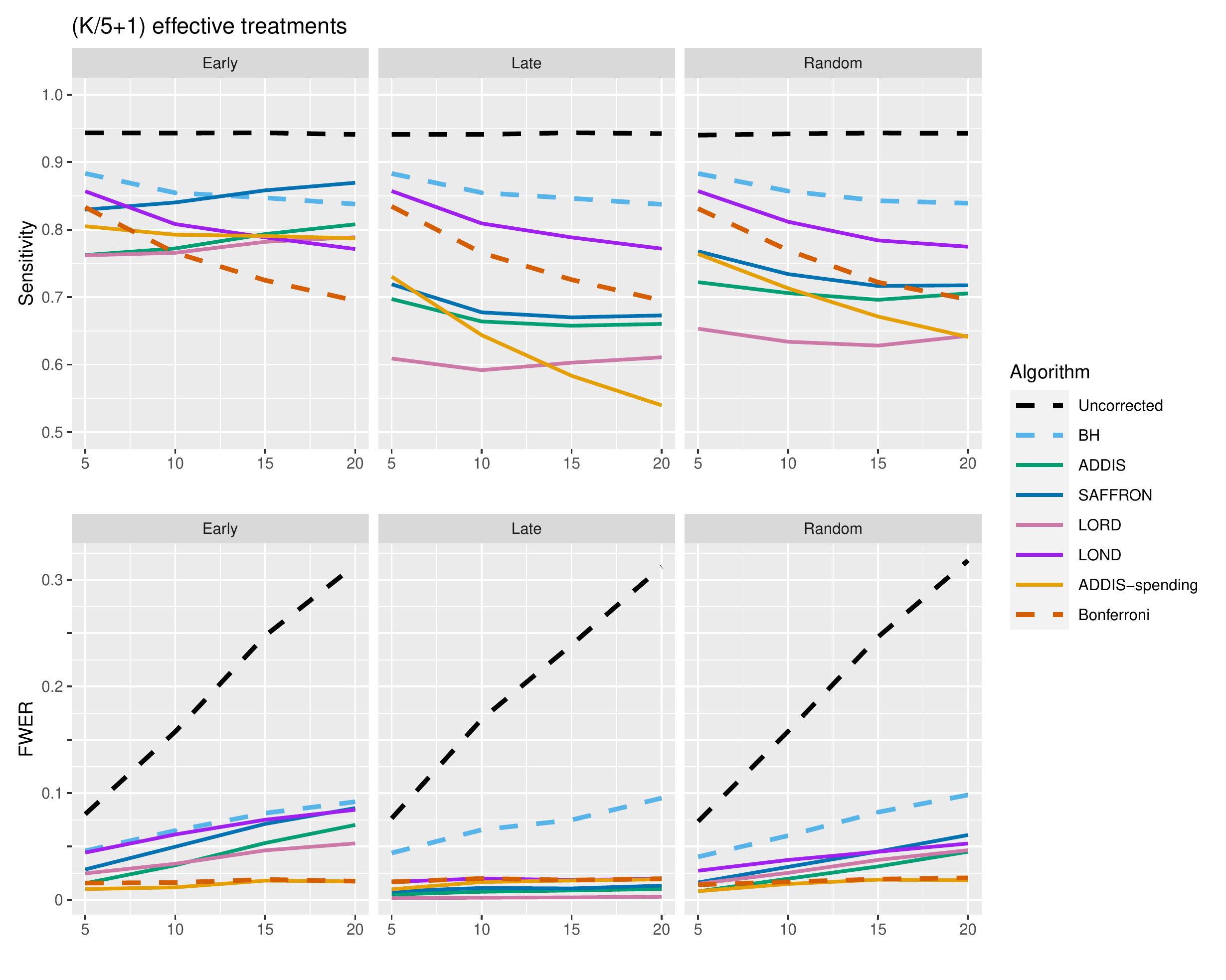}
    \caption{Sensitivity and FWER for fixed means and $(K/5+1)$ effective treatments, with $\Nbound = K$.}
    \label{fig:fully_seq_sens_FWER2}
\end{figure}

\begin{figure}[ht]
    \centering
    \includegraphics[width = \textwidth]{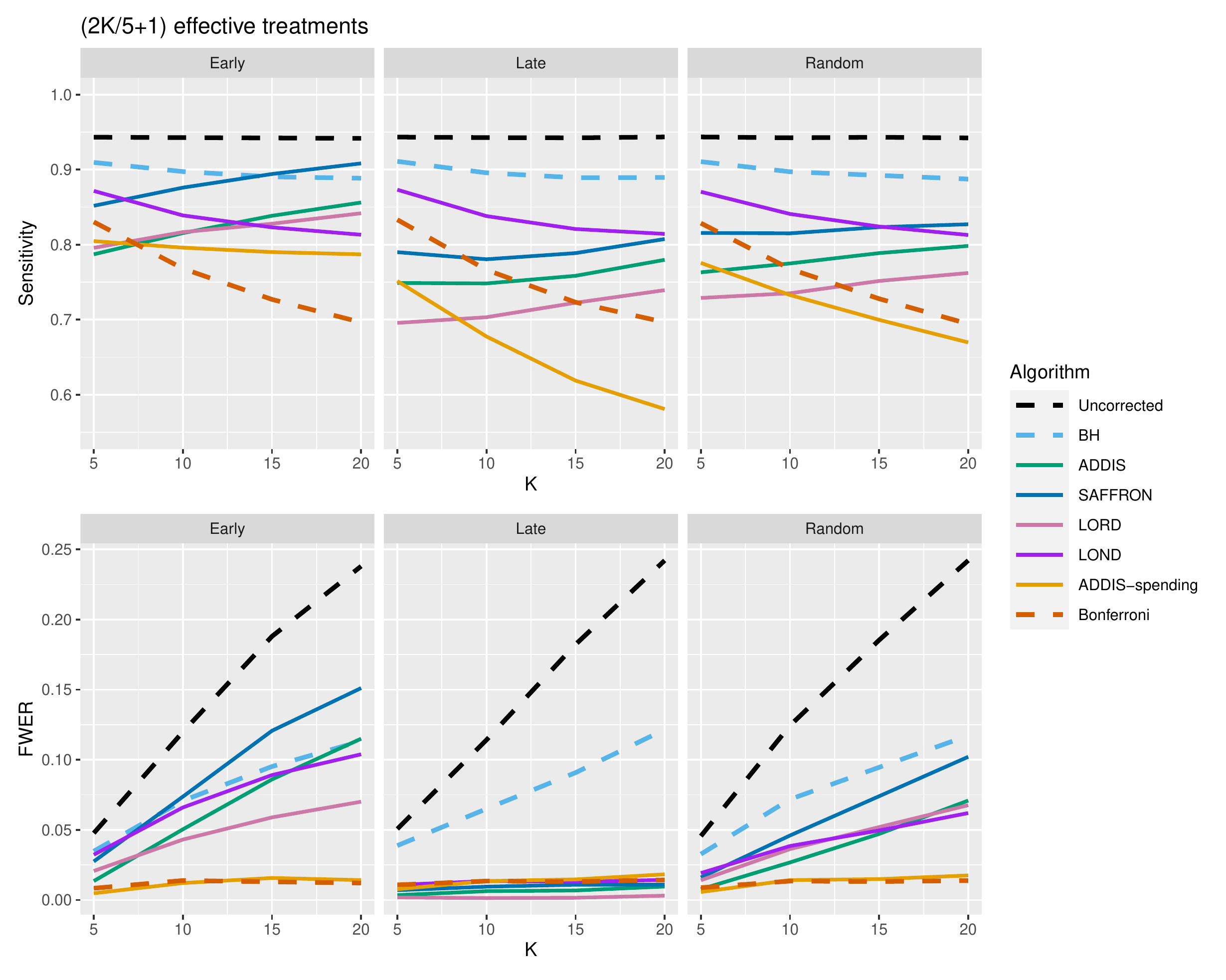}
    \caption{Sensitivity and FWER for fixed means and $(2K/5+1)$ effective treatments, with $\Nbound = K$.}
    \label{fig:fully_seq_sens_FWER3}
\end{figure}

\begin{figure}[ht]
    \centering
    \includegraphics[width = \textwidth]{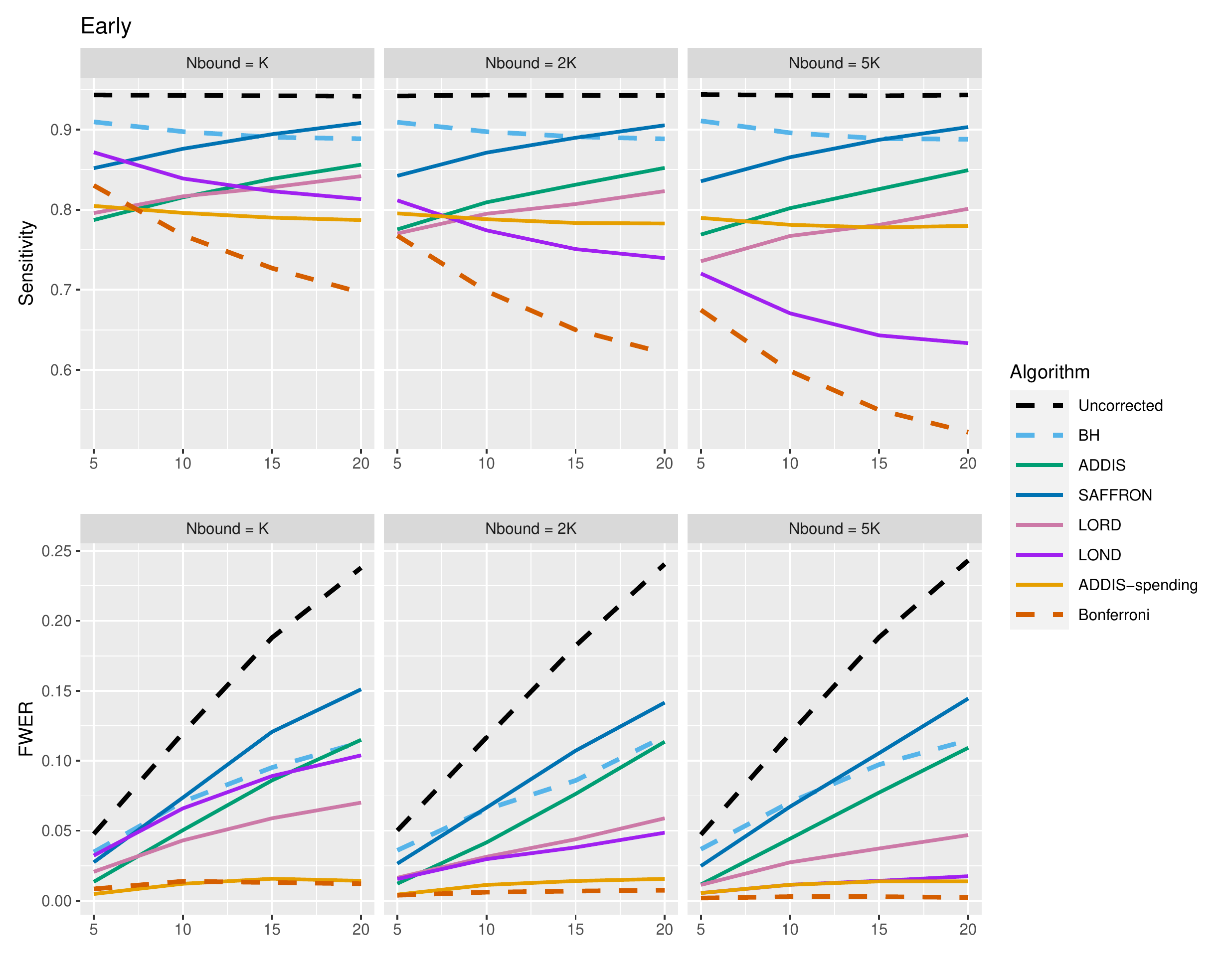}
    \caption{Best case sensitivity with corresponding FWER for fixed means and ($2K/5+1$) effective treatments, with varying $\Nbound$.}
    \label{fig:fully_seq_sens_Nbound1}
\end{figure}

\begin{figure}[ht]
    \centering
    \includegraphics[width = \textwidth]{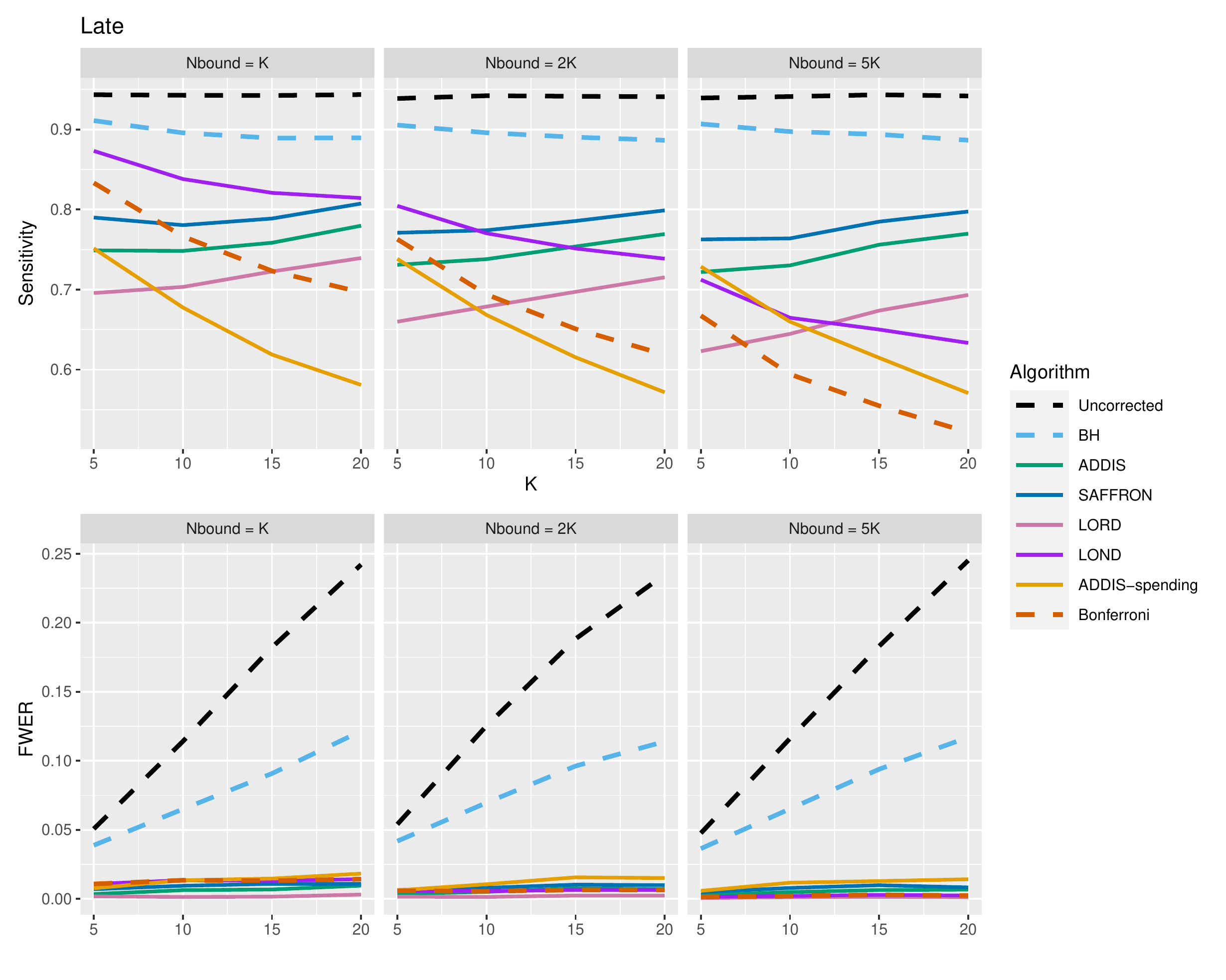}
    \caption{Worst case sensitivity with corresponding FWER for fixed means and ($2K/5+1$) effective treatments, with varying $\Nbound$.}
    \label{fig:fully_seq_sens_Nbound3}
\end{figure}

\begin{figure}[ht]
    \centering
    \includegraphics[width = \textwidth]{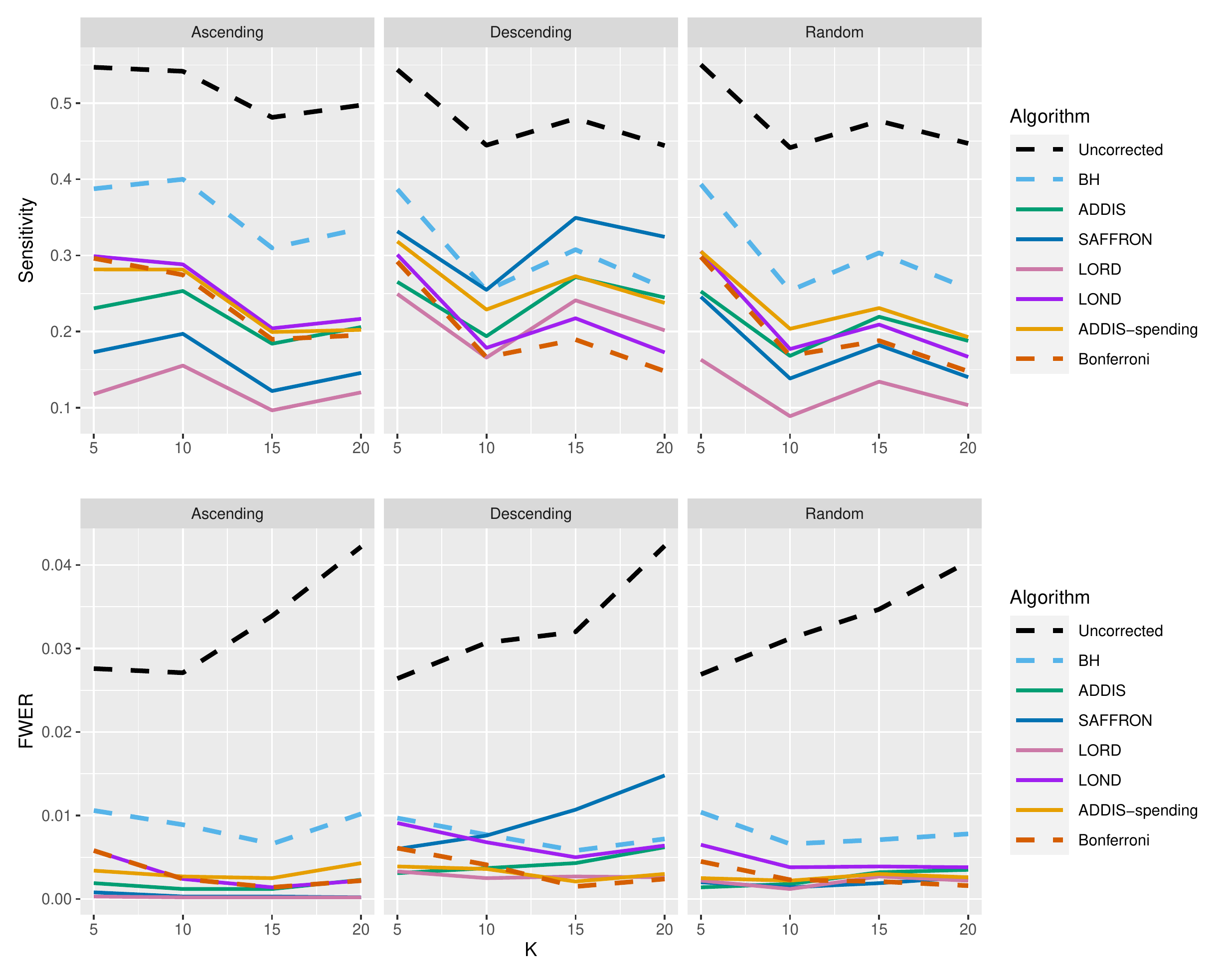}
    \caption{Sensitivity and FWER for the staircase scenarios. Here $\Nbound = 2K$ for the sensitivity plots and $\Nbound = K$ for the FWER plots.}
    \label{fig:staircase_sens}
\end{figure}








\clearpage

\subsection{Batched algorithms}
\label{appendix:batched}

\vspace{12pt}


\rotatebox{90}{\begin{minipage}{0.9\textheight}
\includegraphics[width = \textwidth]{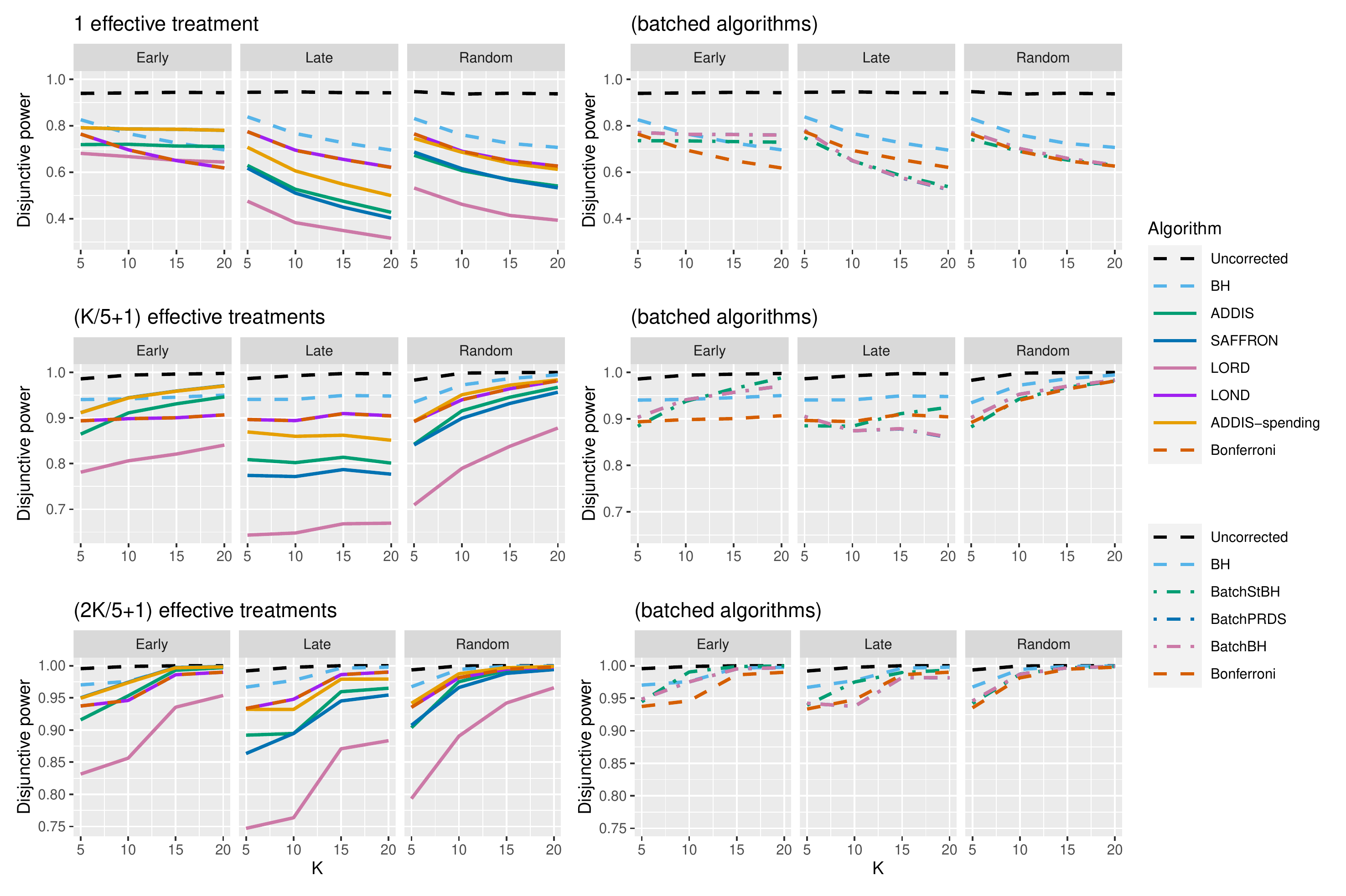}
    \captionof{figure}{Comparison of the disjunctive power for fully sequential and batched online algorithms, with $\Nbound = 2K$.}
    \label{fig:batched_dpower}
\end{minipage}}

\section{Case study: STAMPEDE trial}
\label{appendix:STAMPEDE}

Table~\ref{tab:STAMPEDE_rej2} shows the results if the ordering of treatment arms~B and~C within the first batch were switched. The results are similar as before, except that ADDIS-spending and ADDIS now reject hypotheses~C and~G when $\alpha \in \{0.05, 0.1\}$ and $\alpha = 0.1$, respectively. \\

\begin{table}[ht!]
    \centering
    \begin{tabular}{l | l l l | l l l}
         \textbf{Algorithm} & \multicolumn{3}{c}{\textbf{Hypotheses rejected}} & \multicolumn{3}{|c}{$\bm{\alpha_8}$}\\
         & $\alpha = 0.025$ & $\alpha = 0.05$ & $\alpha = 0.1$ & $\alpha = 0.025$ & $\alpha = 0.05$ & $\alpha = 0.1$ \\ \hline
         Uncorrected & C, E, G & C, E, G & C, E, G & 0.0250 & 0.0500 & 0.1000\\
         Bonferroni & G & G & G & 0.0013 & 0.0025 & 0.0050\\
         ADDIS-spending & G & \textbf{C}, G & \textbf{C}, G & 0.0005 & 0.0011 & 0.0021\\
         BH & C, G & C, G & C, E, G & --  & -- & --\\
         ADDIS & -- & G & \textbf{C}, G & 0.0003 & 0.0016 & 0.0062\\
         SAFFRON & G & C, G & C, E, G & 0.0041 & 0.0165 & 0.0412\\
         LORD & -- & -- & -- & 0.0001 & 0.0002 & 0.0003\\
         LOND & G & G & G & 0.0025 & 0.0050 & 0.0100 \\
         BatchBH & G & C, G & C, E, G & 0.0019 & 0.0057 & 0.0151\\
         BatchPRDS & G & C, G & C, E, G & 0.0019 & 0.0057 & 0.0151\\
         BatchStBH & C, G & C, E, G & C, E, G & 0.0381 & 0.1015 & 0.1238\\
    \end{tabular}
    \caption{Rejections and current significance level $\alpha_8$ of different algorithms using the results of the STAMPEDE trial, with the ordering of treatment arms B and C switched. The new rejections are shown in bold font.}
    \label{tab:STAMPEDE_rej2}
\end{table}